\documentclass{IEEEtran}
\usepackage{cite}
\usepackage{amsmath,amssymb,amsfonts,amsxtra}
\usepackage{algorithmic}
\usepackage{graphicx}
\usepackage{textcomp}
\IEEEoverridecommandlockouts
\usepackage{textcomp}
\usepackage{xcolor}
\usepackage{setspace}
\usepackage{color} 
\usepackage[normalem]{ulem}
\usepackage{subcaption}
\usepackage{tabularx}
\usepackage{array}

\DeclareMathOperator*{\argmin}{argmin} 
\DeclareMathOperator*{\argmax}{argmax} 

\def\BibTeX{{\rm B\kern-.05em{\sc i\kern-.025em b}\kern-.08em
    T\kern-.1667em\lower.7ex\hbox{E}\kern-.125emX}}

\makeatletter
\newcommand{\thickhline}{%
	\noalign {\ifnum 0=`}\fi \hrule height 1pt
	\futurelet \reserved@a \@xhline
}

\begin{document}

\title{A Statistical Approach to Signal Denoising Based on Data-driven Multiscale Representation}

\author{{Khuram Naveed}, {Muhammad Tahir Akhtar, ~\IEEEmembership{Senior Member,~IEEE}, {Muhammad Faisal Siddiqui} and {Naveed ur Rehman}}
\thanks{Manuscript received xx, 20xx; revised xx 20xx.}
\thanks{Copyright (c) 2020 IEEE. Personal use of this material is permitted. However, permission to use this material for any other purposes must be obtained from the IEEE by sending a request to pubs-permissions@ieee.org.}
\thanks{Dr. K. Naveed, Dr. M. F. Siddiqui, and Dr. N. ur Rehman  are with the Department of Electrical Engineering, COMSATS University Islamabad (CUI), Islamabad, Pakistan (emails: khurram.naveed@comsats.edu.pk, faisal\_siddiqui@comsats.edu.pk, naveed.rehman@comsats.edu.pk). }
\thanks{Dr. Akhtar is with the Department of Electrical and Computer Engineering, School of Engineering \& Digital Sciences, Nazarbayev University, Kabanbay Batyr Avenue 53, Nur-Sultan City, Republic of Kazakhstan. (emails:  muhammad.akhtar@nu.edu.kz, akhtar@ieee.org).}
}

\markboth{Preprint}{Naveed et al.: Data-Driven Signal Denoising Based on VMD and CVM Statistic}

\maketitle
	
\begin{abstract}
We develop a data-driven approach for signal denoising that utilizes variational mode decomposition (VMD) algorithm and Cramer Von Misses (CVM) statistic. In comparison with the classical empirical mode decomposition (EMD), VMD enjoys superior mathematical and theoretical framework that makes it robust to noise and mode mixing. These desirable properties of VMD materialize in segregation of a major part of noise into a few final modes while majority of the signal content is distributed among the earlier ones. To exploit this representation for denoising purpose, we propose to estimate the distribution of noise from the predominantly noisy modes and then use it to detect and reject noise from the remaining modes. The proposed approach first selects the predominantly noisy modes using the CVM measure of statistical distance. Next, CVM statistic is used locally on the remaining modes to test how closely the modes fit the estimated noise distribution; the modes that yield closer fit to the noise distribution are rejected (set to zero). Extensive experiments demonstrate the superiority of the proposed method as compared to the state of the art in signal denoising and underscore its utility in practical applications where noise distribution is not known a priori.
\end{abstract}


\section{Introduction}
\label{sec:introduction}
Signals from various practical applications are subject to unwanted noise owing to various physical limitations of acquisition systems, e.g., audio recording systems, lidar systems, EEG and ECG acquisition systems etc. Consequently, to avoid any false decisions based on these noisy signals, it is necessary to remove the unwanted noise beforehand. For this purpose, earlier denoising approaches employed filtering either in time domain or in the transform domain. The filtering methods in original signal domain are referred to as time domain filters which are mostly based on the least mean square (LMS) principle of noise smoothing \cite{feintuch1976LMS,naveed2016AP}. On the other hand, transform domain filters are facilitated by the differentiability of signal and noise in the transform domain \cite{vaseghi2008advanced,mallat1999wavelet}.

The problem of additive white Gaussian noise (wGn) removal has been optimally solved for wide sense stationary signals, i.e., signals with perfectly known invariable statistics, using the Weiner filter. However, that approach may not be adequate in practical settings due to the following reasons. Firstly, majority of real life signals are nonstationary in that their attributes (statistics) change with time. Secondly, the assumed wGn model may not always be used to characterize noise in time series data, e.g., EEG/ECG signals. Consequently, more evolved techniques capable of accounting for the nonstationarity of signal and non-Gaussianity of noise are required to process practical signals. 

Discrete wavelet transform (DWT) is a multiscale method to process the non-stationary signals that exhibits property of sparse distribution of signal singularities within its coefficients. The noise coefficients, on the other hand, have lower amplitudes and uniform spread \cite{mallat1999wavelet}. That allows to differentiate between signal and noise coefficients using a suitable threshold, e.g., universal threshold-based approaches \cite{donoho1995Visu,aminghafari2006MWD} and Steins unbiased risk estimate (Sure)-based approaches \cite{donoho1995Sure,blu2007Surelet}. Similarly, shrinkage functions based on the probability distribution of signal and noise coefficients are also derived using Bayesian estimators, e.g., \cite{chang2000BayesShrink,abramovich1998BayesianPosteriorMedians}.

The above-mentioned methods require a prior information about the signal and noise (distribution) models in order to estimate the threshold or derive the shrinkage (thresholding) function to suppress the noise. A variety of noise models are available based on the experimental studies \cite{vaseghi2008advanced}, however, these models do not fully account for the factors contributing to the noise during acquisition. Consequently, noise is abstractly modeled using these experimental models within the denoising methods. A more challenging task in this regard involves the specification of a generalized signal model owing to the arbitrary nature of information generally found within the times series data. Secondly, specification of prior models restricts the efficacy of these methods in real world signals.

This issue has been partially addressed in framework proposed in \cite{ur2017DWTGoF} which combines DWT with the goodness of fit (GoF) test. Hereafter, this approach is called as DWT-GoF method. It is worth mentioning that the DWT-GoF method requires only a prior noise model. Here, noise is expediently modeled as a zero-mean additive wGn that is conventionally used to model the random noise in the data-acquisition and communication systems, for example. The detection of wGn at multiple wavelet scales is facilitated by the fact that the DWT preserves the Gaussianity of noise. This essentially requires detection and rejection of wavelet coefficients fitting the Gaussian distribution for denoising. Henceforth, the DWT-GoF method \cite{ur2017DWTGoF} rejects noise from DWT scales by estimating the GoF of Gaussian distribution on the multiscale coefficients. An improved version of the DWT-GoF method has been proposed in \cite{naveed2018DTCWTGoF,naveed2017DTCWTGoFConference} that employ GoF test along with the dual tree complex wavelet transform (DTCWT), which is called the DT-GOF-NeighFilt method in the sequel. The key feature of the DT-GOF-NeighFilt method is to incorporate a novel neighborhood filtering technique to minimize the loss of signal details while rejecting the noise. Apart from the GoF test, other hypothesis testing tools such as False discovery rate (FDR), Bayesian local false discovery rate (BLFDR) are also used in combination with wavelet transforms for signal denoising \cite{abramovich1996FDR2,lavrik2008BLFDR}.

Another avenue for multiscale denoising involves data-driven decomposition techniques. For instance, empirical mode decomposition (EMD) \cite{huang1998EMD} that employs a data-driven approach to extract principal oscillatory modes from a signal. Within EMD, local extrema (maxima/minima) of a signal are interpolated to obtain its upper and lower envelops and their mean is subtracted from the original signal. This process, called sifting, continues recursively until zero-mean oscillatory components, namely intrinsic mode functions (IMFs), are obtained. Owing to this ability to expand a signal into its IMFs, EMD is considered well suited for processing the non-stationary signals generally encountered in practice. Keeping in view its efficacy for 1D signals, several variants of EMD have also emerged for multichannel signals, e.g., multivariate EMD (MEMD) \cite{rehman2009MEMD}, dynamically sampled MEMD \cite{rehman2015DS-MEMD}, etc.

When employed for denoising, EMD aims at detecting the IMFs representing the (oscillatory) signal parts and rejecting the IMFs corresponding to the non-oscillatory noise. A wavelet-inspired interval-thresholding function is used for detecting the oscillatory signal parts from the noisy IMFs \cite{kopsinis2009EMD-IT}. Specifically, the EMD-based interval thresholding (EMD-IT) \cite{kopsinis2009EMD-IT} aims to detect the oscillations separated by two consecutive zero crossings. This is achieved by comparing the extrema of an interval against a threshold value leading to either retention or rejection of the whole interval. The interval thresholding has since been used within a variety of denoising methods and has seen several variants including interval thresholding based on histogram partition \cite{chen2008EMDIThistogram}, MEMD-based interval thresholding \cite{Hao2017} and a purely multivariate interval thresholding \cite{ur2019MEMD-MIT}.

Instead of performing thresholding, the work in \cite{flandrin2004detrending,mert2014DFA} employed statistical tools to detect
the relevant (signal) modes for a partial reconstruction of the denoised signal.
However, these denoising approaches may result in suboptimal performance due to the mode mixing (i.e., manifestation of multiple IMFs within a single IMF) property of EMD and its sensitivity to noise and sampling. Essentially, the aforementioned shortcomings within EMD framework result in leakage of noise into a few signal modes which leads to their rejection resulting in suboptimal denoising. The lack of mathematical foundation of the EMD limits the chances of rectification of these issues within its framework. The issue of noise presence within the selected relevant IMFs was better handled by partial reconstruction of the thresholded relevant modes \cite{yang2015emdparTh}.

The recently proposed variational mode decomposition (VMD) is based on optimization of a variational problem to obtain an ensemble of a fixed number of band limited IMFs (BLIMFs) \cite{dragomiretskiy2014VMD}. Owing to its sound mathematical foundation, VMD successfully avoids mode mixing and is robust to noise and sampling unlike EMD \cite{dragomiretskiy2014VMD,ali2018hybrid}. 
From the view point of denoising, a very important feature of VMD is its ability to segregate the desired signal into a few initial BLIMFs while noise is mostly stashed into a few final BLIMFs. Hence, by rejecting the modes with noise, a good estimate of the true signal may be obtained by partial reconstruction.

A literature review shows that the existing VMD-based denoising approaches select relevant signal modes by comparing the probability distribution function (PDF) of an individual BLIMF against the PDF of the noisy signal. This is well founded because a distribution function is generally reflective of the signal present within the noisy data. An estimate of the signal present in a BLIMF may be obtained by measuring the closeness of its PDF with that of the noisy signal, for example, by employing Euclidean distance \cite{ren2017ED-VMD}, Bhatacharya distance \cite{li2019BhataCharyaVMD}, etc. Therein, the modes statistically close to the noisy signal are retained as relevant signal modes while largely dissimilar modes are rejected as noise. For a detailed study on the efficacy of various statistical distances for estimating relevant modes, the interested reader is referred to \cite{ma2017VMDHurdoffs}. The result presented in \cite{ma2017VMDHurdoffs} show that the Hausdoffs distance \cite{Hurdoffs} yields the best denoising performance. Apart from that, the method in \cite{liu2016VMDDFA} selects relevant modes using the detrended fluctuation analysis (DFA) (originally used with EMD within the EMD-DFA method \cite{mert2014DFA}) that estimates the randomness of data by observing the lack of trend. This method, hereafter referred as VMD-DFA \cite{liu2016VMDDFA}, rejects the noisy BLIMFs and reconstructs the denoised signal based on the remaining modes.

In this paper, we present a novel approach to signal denoising that uses Cramer Von Misses (CVM) statistic locally on multiscale signal decomposition obtained through VMD. A nonlinear thresholding scheme based on Goodness of Fit (GoF) test is utilized to test whether the obtained CVM values (at multiple scales) conform to noise distribution or not. Those parts of the signal which conform to noise are discarded while the rest are retained. Our approach is different and more effective than other denoising methods, e.g., [27], owing to the inherent robustness of CVM statistic in testing for a given data distribution; we refer readers to a detailed description of empirical distribution function (EDF) based statistics, including CVM, for detecting normality \cite{stephens1974edf}.     
Specifically, we propose a robust multistage procedure whereby first the predominantly noisy modes are detected using the CVM distance which are subsequently used to estimate the noise distribution. Finally, an empirical GoF test based on CVM statistic and the estimated distribution of noise is used to reject the noise coefficients from within the remaining modes. The main contributions of this work include:
\begin{itemize}
  \item Estimation of noise distribution model from within the noisy signal that is facilitated by the effective segregation of noise and true-signal by the VMD into separate groups of modes owing to its robustness to noise and mode mixing.
  \item The use of the robust CVM distance based on EDF statistic as a means to detect relevant signal modes and the same time reject the predominantly noise modes.
  \item The annihilation of noise from within the remaining relevant signal modes by estimating how closely the estimated the noise distribution fits the local segments of the selected IMFs using the CVM test.
\end{itemize}
To validate the performance our method, extensive computer simulations have been carried out for denoising a variety of benchmark signals corrupted by artificially generated Gaussian noise. Furthermore, the efficacy of the proposed method is demonstrated by denoising a few (real) EEG signals corrupted by actual (non-Gaussian) sensor noise. 

The rest of paper is organized as follows: Section \ref{prelim} provides the preliminaries related to the proposed methodology that is subsequently presented in Section \ref{proposd}. Section \ref{results} reports experiments analyzing the performance of our proposed work while Section \ref{EEGresults} presents a few practical denoising examples. Finally, conclusion along with the future prospects of this work are discussed in Section \ref{conclusion}.

\begin{figure*}[t]
	\begin{center} \includegraphics[scale=0.55]{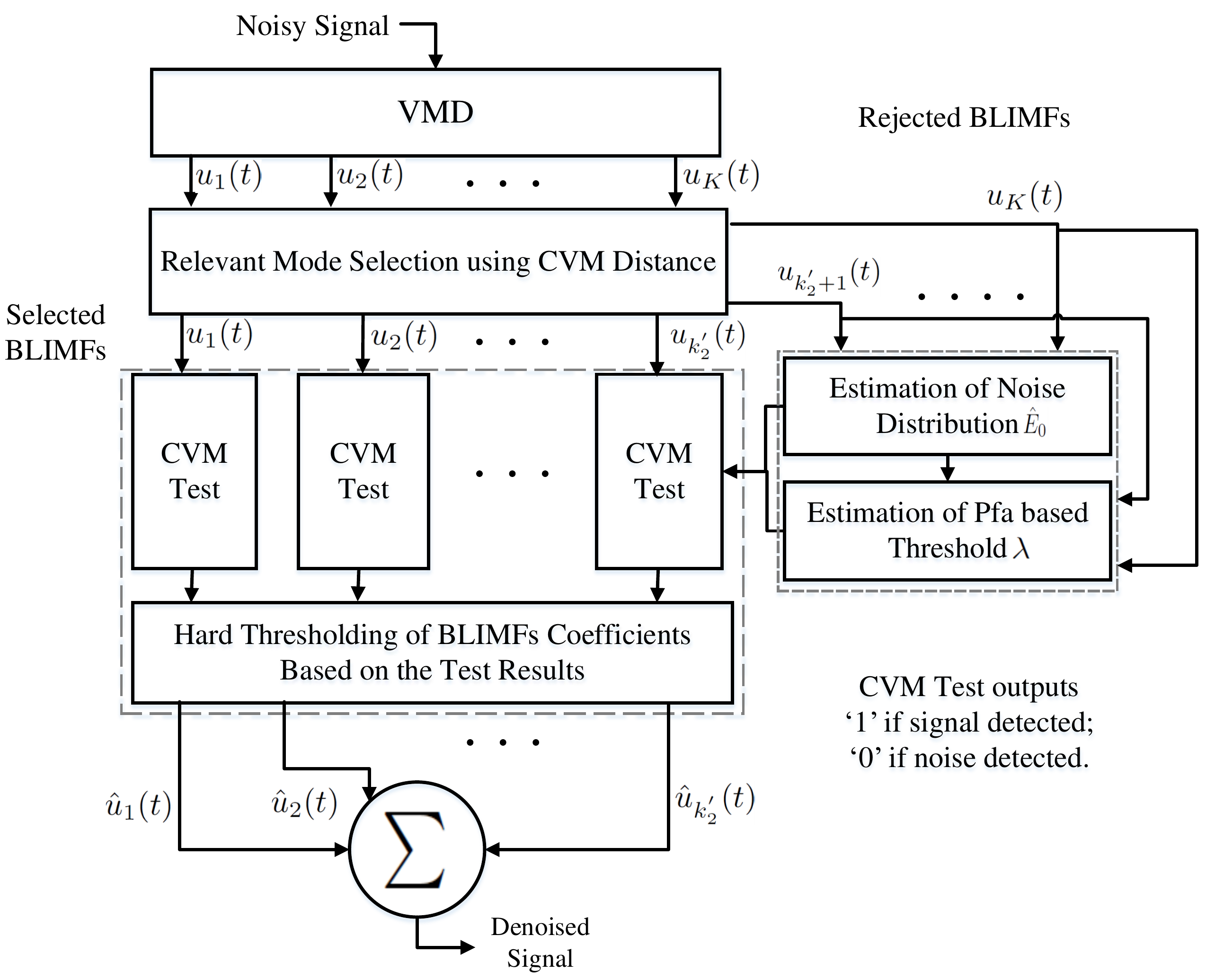} \end{center}
	\caption{Block diagram of the proposed denoising method.}
	\label{Fig:01}
\end{figure*}
\section{Preliminaries}
\label{prelim}

\subsection{Variational Mode Decomposition (VMD)}
VMD employs an entirely non-recursive approach to decompose a signal ${y}(t),\ \forall \ t=1,\ldots,N$, into $K$ predefined modes BLIMFs ${u}_k(t),\ \forall \ t=1,\ldots,N$. This is achieved by first finding the center frequencies $w_k$ and then an ensemble of compact BLIMF by solving the following constrained variational problem \cite{dragomiretskiy2014VMD}
\begin{align}\label{Eq:01}
    \argmin\limits_{\{u_k,w_k\}} \sum_{k=1}^K\left \| \partial_t \left [ \left ( \delta (t) + \frac{j}{\pi t} \right ) \ast {u}_k(t) \right ]e^{-jw_k t} \right \|,
\end{align}
subject to
\begin{align}\label{Eq:02}
    y(t) = \sum_{k=1}^K u_k(t),
\end{align}
where $\delta(t)$ denotes the Dirac distribution, and $\ast$ represents the linear convolution. In order to enforce the constraint that the aggregate of the total number of $K$ modes $u_k(t)$, where $k=1,\ldots,K$, amount to the original signal; Lagrangian multipliers $\gamma(t)$ are used
\begin{align}\label{Eq:03}\nonumber
         \argmin\limits_{\{u_k,w_k\}} & \sum_{k=1}^K\left \| \partial_t \left [ \left ( \delta (t) + \frac{j}{\pi t} \right )\ast {u}_k(t) \right ]e^{-jw_k t} \right \|\\
                & + \Big\|\sum_{k=1}^K u_k - {y}(t)\Big\|^2 + \displaystyle \Big<\gamma(t), \sum_{k=1}^K {u}_k - {y}(t)\Big>,
\end{align}
where the quadratic data fidelity term $\|\sum_{k=1}^K {u}_k(t) - {y}(t)\|^2$ is used for its accelerated convergence and to ensure minimum squared error \cite{dragomiretskiy2014VMD}. This way, the center frequencies $w_k$ required to find compact modes that successfully avoid mode mixing are estimated by solving (3) using the alternating direction method of multiplied (ADMM) \cite{dragomiretskiy2014VMD}. 
Further details of the algorithm can be found in \cite{dragomiretskiy2014VMD}.

\subsection{Cramer Von Mises (CVM) statistics}
CVM statistic \cite{cramer1928CVMtest} belongs to a class of statistical distances \cite{stephens1974edf} that estimate how closely a dataset or observations follow a given distribution function. In this regard, the CVM statistic requires an estimate of the distribution of given observations, that is obtained using the EDF. It is worth mentioning that EDF happens to be a robust model of distribution even for small-sized data and is easy to compute \cite{d2017goodness}. More importantly, EDF is a discrete approximation of the cumulative distribution function (CDF) that means a distribution test is realized by testing how close an EDF of the data at hand is from the CDF of that reference distribution. This type of testing framework is termed as GoF test of a distribution on given dataset whereby EDF-based distances, e.g., Kolmogrov Smirnov (KS) statistic \cite{smirnov1948KS}, Anderson Darling (AD) statistic \cite{anderson1954ADtest}, CVM statistic etc., are used to estimate the measure of fit of the reference CDF on the EDF of data at hand.

Given the CDF $E_0(z)$ corresponding to reference distribution and EDF $E(z)$ of given observations, the CVM statistic is given as follows
\begin{align}\label{Eq:04}
    \Delta = \int_{-\infty}^{\infty} \Big(E_0(z)-E(z)\Big)^2 d(E_0(z)),
\end{align}
where $z$ denotes the support of the distribution function. Note that \eqref{Eq:04} involves an indefinite integration and is practically not computable. Therefore, its computable numerical adaptation is presented by D'Augustino in \cite{d2017goodness}
\begin{align}\label{Eq:05}
    \Delta = \frac{1}{12L} + \sum_{t=1}^{L}{\left (E_0\Big(z^{'}(t)\Big)-\frac{(2t-1)}{L}\right )},
\end{align}
where $z^{'}(t)$ denotes a set of observations having finite length $L$, i.e., $t=1,2,\cdots,L$.

The GoF test based on CVM statistic is realized by estimating the significance level or threshold $\lambda$ that specifies the maximum value of the test statistic $\Delta$ \eqref{Eq:05} that is sufficient to suggest a {\emph{close-fit}}. The GoF testing framework checks the following binary hypothesis:
\begin{align}\label{Eq:06}\nonumber
    & \mathcal{H}_0 : \Delta \leq \lambda, \\
    & \mathcal{H}_1 : \Delta > \lambda,
\end{align}
where $\mathcal{H}_0$ denotes the null hypothesis suggesting a {\emph{close-fit}} of null (or reference) distribution on the given data while the $\mathcal{H}_1$ denotes the alternate hypothesis of {\emph{no-fit}}. The threshold parameter $\lambda$ is estimated by minimizing the probability of false alarm ($P_{\textrm{fa}}$), i.e., false detection rate of the alternate hypothesis $\mathcal{H}_1$ given the null hypothesis $\mathcal{H}_0$, and is mathematically stated as follows
\begin{align}\label{Eq:07}
    P_{\textrm{fa}} = \textrm{Prob}( \mathcal{H}_1| \mathcal{H}_0) = \textrm{Prob}( \Delta > \lambda| \mathcal{H}_0),
\end{align}
where $\textrm{Prob}(\cdot)$ denotes the probability of the event stated within the parenthesis. Here, the $P_{\textrm{fa}}=\alpha$, where $\alpha$ is kept very small, e.g., $\alpha=10^{-2}-10^{-4}$ to minimize the false detection of noise as signal \cite{ur2017DWTGoF, lei2011SS1,wang2009SS2}.
\begin{figure}[t]
	\begin{minipage}[b]{1\linewidth} \centerline{\includegraphics[scale=0.35]{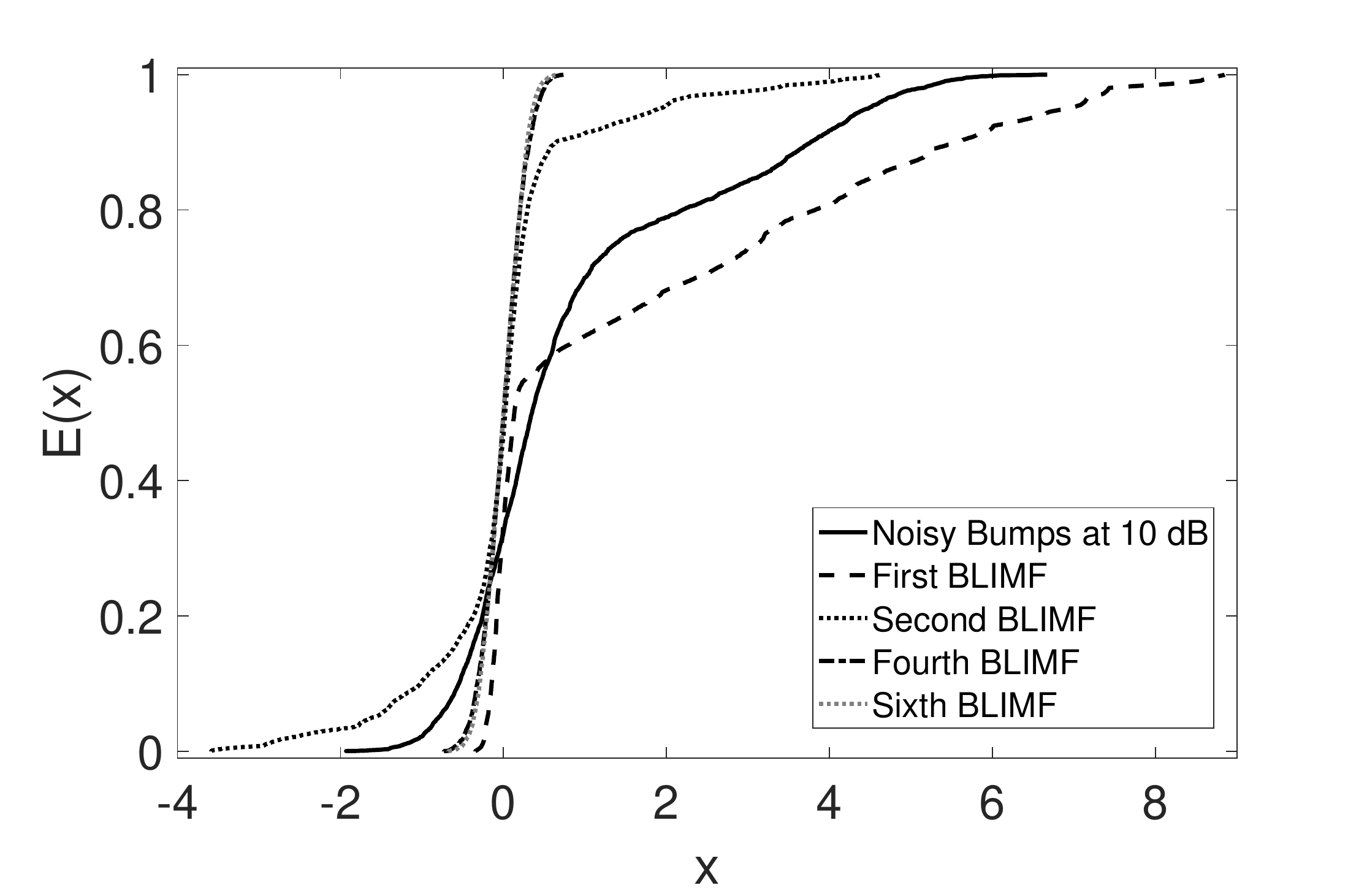}} \centerline{(a)} \end{minipage}
	\begin{minipage}[b]{1\linewidth} \centerline{\includegraphics[scale=0.35]{Fig02b.pdf}} \centerline{(b)} \end{minipage}
	\caption{Depiction of how CVM statistic estimates the closeness of the noisy signal from its BLIMFs. That is demonstrated by plotting EDFs of the noisy Bumps and Blocks signal along with EDF of their (selected) BLIMFs in (a) and (b) respectively.}
	\label{Fig:02}
\end{figure}
\section{Description of proposed approach}
\label{proposd}
Consider the signal model
\begin{align}\label{Eq:08}
    y(t) = x(t) + \psi(t),
\end{align}
where $y(t)$, $x(t)$ and $\psi(t)$ denote the noisy, true signal, and additive noise component, each of length $N$. It is customary to assume $\psi(t)$ being modeled as $\mathcal{N}(0,\sigma)$, i.e., zero-mean wGn process with variance $\sigma^2$. However, the additive noise in the real-life signals may be non-Gaussian. Therefore, the existing denoising approaches developed on the assumption of wGn may have a limited scope for the practical signals. 
In this work, we propose to address the noise removal in practical signals using a robust multi step procedure based on the VMD and CVM statistic. The main idea of the proposed work is to estimate noise distribution from the dominantly noisy VMD modes of the noisy signal and then use it as a mean to detect the noise coefficients from the rest of the modes using the CVM statistic.

As stated earlier VMD effectively segregates the true signal from noise whereby signal details are mostly concentrated in a few initial BLIMFs. This is because of the well posed variational problem in \eqref{Eq:01} that leads to the expansion of the noisy signal $y(t)$ as an ensemble of a set of $K$ BLIMFs $\{u_k(t), \ \forall \ k=1,\ldots,K\}$ as given in (2). Given that $1<k_1^{'}<k_2^{'}<K$, these BLIMFs may be largely categorized into following categories owing to the robust architecture within VMD to segregate signal and noise \cite{ma2017VMDHurdoffs,liu2016VMDDFA}:
\begin{itemize}
	\item Initial modes $\{u_k(t), \ k<k_1^{'}\}$ are composed of mostly signal content;
	\item Intermediate modes $\{u_k(t), \ k_1^{'}<k<k_2^{'}\}$ mostly contain signal plus noise;
	\item Final modes $\{u_k(t), \ k>k_2^{'}\}$ composed of predominantly noise.
\end{itemize}

The conventional VMD-based denoising approaches exploit this representation to perform the signal denoising whereby the modes with dominant signal, i.e., $\{u_k(t), \ k<k_1^{'}\}$, are employed as relevant modes for a partial reconstruction of the denoised signal. The rest of the modes are simply rejected. This methodology results in a significant loss of the desired signal-details due to the rejection of intermediate (signal plus noise) modes, i.e., $u_k(t)$ for $k_1^{'}<k<k_2^{'}$, along with the modes with dominant noise, i. e., $\{u_k(t)$ for $k>k_2^{'}\}$.

In order to maximally preserve the desired signal information, the proposed framework only rejects the dominantly noise modes while the remaining modes are preserved as relevant signal modes using the CVM statistic. Hence, our definition of relevant modes includes the initial modes containing mostly signal and intermediate modes composed of both signal and noise. Subsequently, the selected relevant BLIMFs are cleansed of noise via a statistical thresholding function that operates by first estimating the noise distribution from the rejected noise modes which is then used to detect noise coefficients from selected modes using the CVM test. Finally, the denoised signal is partially reconstructed using the thresholded BLIMFs. This robust multistage procedure is depicted using the block diagram in Fig. \ref{Fig:01} where each stage is explained in detail in the subsequent sections.

\subsection{Relevant Mode Selection}
This section details the process adopted to select BLIMFs containing signal information with or without noise, i.e., the relevant modes $\{u_k(t), \ k<k_2^{'}\}$. This is achieved by detecting the BLIMFs entirely composed of noise, i.e., $\{u_k(t), \ k>k_2^{'}\}$. In this regard, CVM distance based on EDF statistics is used to estimate the statistical distance between the noisy signal and the individual BLIMF.

\subsubsection{Rationale}
To identify relevant signal modes, conventional VMD denoising approaches investigate signal content in a BLIMF by computing some distance measure $D_k^{'}$ between the empirical PDFs of the noisy signal and the BLIMFs, as given below
\begin{align}\label{Eq:09}
    D_k^{'} = \textrm{Distance}\{p_y,p_{u_k}\},
\end{align}
where $p_y$ and $p_{u_k}$ respectively denote the empirical PDFs of the noisy signal $y(t)$ and the $k$th BLIMF $u_k(t)$. The empirical PDFs $p_y$ and $p_{u_k}$ are estimated by dividing the data in hand into a finite number of bins leading to the construction of a PDF for these bins, e.g., use of KS-density function in \cite{ma2017VMDHurdoffs,li2019BhataCharyaVMD,ren2017ED-VMD}. The issue with this approach is that all the data elements in a bin are assigned the same probability as the probability of the bin where it resides. This compromises the individuality of the data points within the bin resulting in a less robust estimate of the distribution especially for small-sized data.

A robust estimate may be obtained by using the EDF \eqref{Eq:05} which is discrete approximation of the CDF of the data distribution. This makes EDF a robust estimator of data distribution even for small-sized data. Consequently, EDF is frequently used within the GoF-based hypothesis testing in various practical applications \cite{lei2011SS1,naveed2019PoisDen}. Therefore, we propose to use EDF-based distance to obtain a robust estimate of the distance between the noisy signal $y(t)$ and its BLIMFs $u_k(t)$
\begin{align}\label{D1} 
    D_k = \textrm{Distance}\{E_{y}(z),E_{u_k}(z)\},
\end{align}
where $E_{y}(z)$ and $E_{u_k}(z)$ respectively denote EDFs of the noisy signal and the $k$th BLIMF $u_k(t)$.

\begin{figure}[t]
	\begin{center} \includegraphics[scale=0.35]{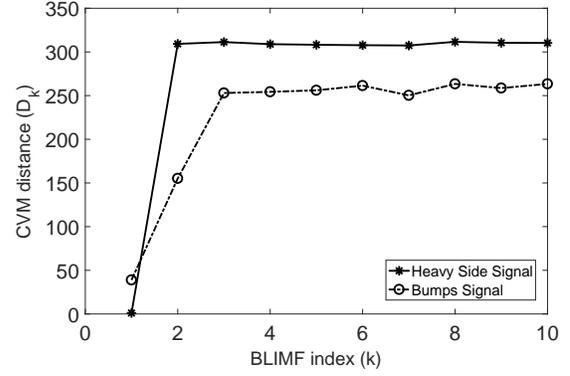} \end{center}
	\caption{An illustration of variation of CVM distance $D_k$ for index $k$ of the BLIMFs of a noisy `Bumps' signal (dotted line) and `Heavy Sine' signal (solid line).}
	\label{Fig:03}
\end{figure}
\subsubsection{Estimating $D_k$ using CVM Statistic}
In order to obtain an estimate $\hat{D}_k$ of the actual (statistical) distances $D_k$ between the $k$th mode $u_k(t)$ and the noisy signal $y(t)$, both of size $N$, we use the CVM statistic as follows
\begin{align}\label{CVMy} 
    \hat{D}_k = \frac{1}{12N} + \sum_{t=1}^{N}{\left (\hat{E}_{y}\Big(u_k(t)\Big)-\frac{(2t-1)}{N}\right )},
\end{align}
where an estimate of the signal EDF $\hat{E}_{y}(z)$ is computed from the noisy signal $y(t)$ though
\begin{align}\label{Eq:12}
    \hat{E}(z) = \frac{1}{N} \sum_{t=1}^{N} \Big(y(t)\leq z\Big).
\end{align}
Here, $z$ denotes the support of the distribution function and $t$ denotes the time index of the data values. The operation $(y(t)\leq z)$ results into a binary decision (i.e., $0$ or $1$) at every index $t$ of the summation whereby, for a given $z$; number of values of $y(t)$ less then or equal to $z$ are accumulated.

In order to develop an insight on how CVM statistic estimates the distance between the noisy signal and the BLIMFs, consider Fig. \ref{Fig:02} which plots EDFs of a few selected BLIMFs with the estimated reference EDF $\hat{E}_{{y}}(z)$ from the noisy signal. These results are obtained for the benchmark signals `Bumps' and `Blocks' (shown later when we present the detailed simulation results). It is observed in Fig. \ref{Fig:02}, that the EDFs of first and second BLIMFs are closest to the EDF of noisy Bumps and Blocks signals that essentially means these initial modes are mostly signal. Contrarily, the EDFs of fourth and sixth BLIMFs are further from the reference EDF which means these higher modes have lesser signal content and more noise.

\subsubsection{Criteria for Selection of Relevant Modes}
The above-detailed discussion indicates that the relevant signal modes may be selected by evaluating slopes of the distances between the consecutive BLIMFs \cite{ma2017VMDHurdoffs,liu2016VMDDFA}. Naturally, significant change in slope between the two adjacent BLIMFs means rapid decline of signal content when moving from earlier to the latter. This ensures that signal will decline further in the forthcoming BLIMFs with the increase in noise. Consequently, the existing methods \cite{ma2017VMDHurdoffs,liu2016VMDDFA} employ maximum slope in the distance-curve to determine a threshold, $k_1^{'}$, to select the relevant modes containing signal details
\begin{align}\label{k1dash} 
    k_1^{'} = \argmax\limits_{k} \left (S_k,S_{k+1},\cdots,S_K \right),
\end{align}
where $S_k$ denotes the slope of the distances of $k$th and $(k+1)$th BLIMFs, computed via
\begin{align}\label{Eq:14}
S_k = |D_{k+1}-D_k|.
\end{align}

Consider Fig. \ref{Fig:03} which plots CVM distances of the modes of two benchmark (noisy) signals `Bumps' and `Heavy Sine' (shown later when we present the detailed simulation results). It is seen from Fig. \ref{Fig:03}, that the CVM distances $D_k$ corresponding to the `Heavy Sine' signal  show maximum slope between the first and second BLIMFs, and declines massively when moving to the second BLIMF and the subsequent ones. This essentially means that most of signal content is concentrated in the first BLIMF. A similar observation can be made for the `Bumps' signal. It is seen that the maximum slope is observed between the second and third BLIMF, and decreases rapidly in the latter modes. This shows that the  signal content is largely concentrated in first two BLIMFs.

The above-detailed procedure, however, selects only dominantly signal modes as the relevant ones $\{u_k(t), \ k<k_1^{'}\}$ for partial reconstruction of the denoised signal. By this definition, the rejected noise modes $\{u_k(t), \ k>k_1^{'}\}$ include the intermediate signal plus noise modes $\{u_k(t), \ k_1^{'}<k<k_2^{'}\}$ that causes loss of signal details. To address this issue, we suggest rejection of only purely noise modes $\{u_k(t), \ k>k_2^{'}\}$ and retention of all the modes containing signal (with or without noise) $\{u_k(t), \ k<k_2^{'}\}$.
The selection of relevant modes according to new definition, i.e., modes containing signal (with or without noise) $\{u_k(t), \ k<k_2^{'}\}$, requires estimation of mode index $k_2^{'}$ that indicates the start of purely noise modes.

For this purpose, we alter the criteria discussed above by dividing the CVM distance curve (plotted in Fig. \ref{Fig:03} for instance) into transient and stable regions where earlier relates to the purely signal modes while latter relates to the modes with noise.
It is observed that the CVM distance plotted in Fig. \ref{Fig:03} have transient phase before the maximum slope and the region after that can be categorized as the stable phase for all the BLIMFs. Since, the transient phase ends with maximum slope that can be seen as the threshold for selecting the signal only modes while an estimate of modes containing signal plus noise may be obtained from the stable phase.
We suggest looking for maximum slope within the stable region which indicates the point of maximum change between partially noisy modes to purely noisy modes. Understandably, maximum slope in the stable region separates the purely noise modes and the modes with signal. Based on the above discussion, maximum slope in the stable region may be obtained as follows
\begin{align}\label{Eq:15}
    k_2^{'} = \textrm{max} \left (S_k,S_{k+1},\cdots,S_K \right), \ \ \ k = k_1^{'}+\cdots,K
\end{align}
where $k_1^{'}$ is obtained from \eqref{k1dash} and $k_2^{'}$ denotes the index of the mode that is followed by noise only modes. Mathematically, the thresholding criteria for selection of relevant modes and rejection noise only modes is then given below
\begin{align}\label{Eq:16}
    \begin{cases}
        \{u_k(t), \ \ k\leq k_2^{'} \} \ \in \ \text{Relevant modes with signal}, \\
        \{u_k(t), \ \ k >k_2^{'} \} \ \in \ \text{Rejected modes with noise}.
    \end{cases}
\end{align}
%
\begin{figure*}[t]
	\begin{center} \includegraphics[scale=0.75]{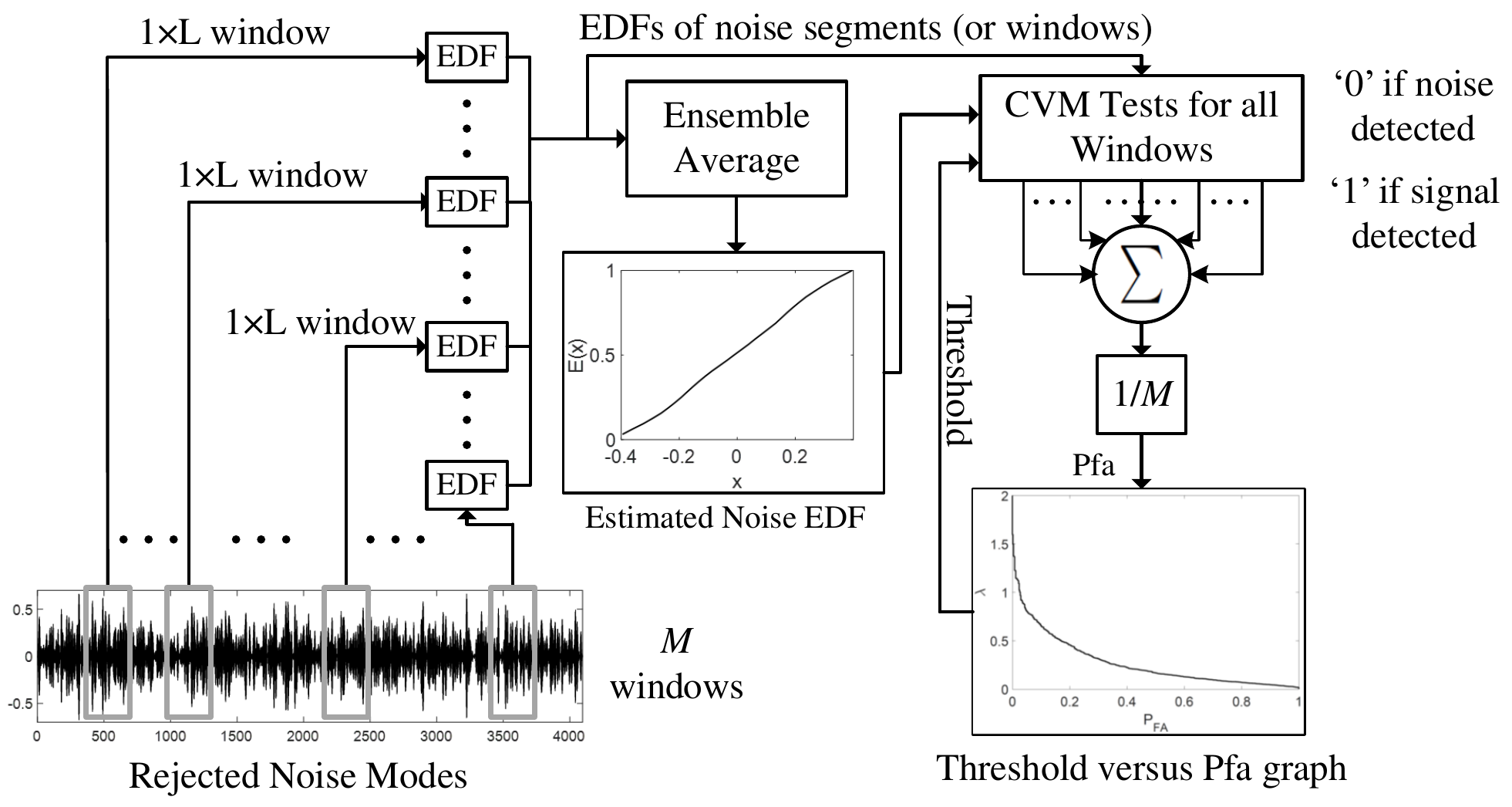} \end{center}
	\caption{Depiction of the procedure adopted for estimation of noise EDF and threshold selection.}
	\label{Fig:05}
\end{figure*}
\subsection{Estimation of Noise Distribution from the Rejected Modes}
In this step, the distribution model governing noise within the VMD BLIMFs is estimated empirically from the rejected modes $\{u_k(t), \ k>k_2^{'}\}$ which are mostly composed of noise. This step of the estimation of noise EDF is an essential part of the proposed approach as indicated by block `CVM Test' in Fig. 1. 
The process adopted for empirical estimation of the noise distribution is illustrated in detail in Fig. \ref{Fig:05} whereby, given a large sized dataset from an unknown distribution; a good estimate of its CDF may be empirically obtained using the ensemble average of the EDF of its local segments. That is a standard procedure used by statisticians to empirically estimate the unknown distribution function governing a given dataset \cite{mcassey2013empiricalMGOF,ur2016multi,naveed2019multiscale}.

Let $u_k^{(n)}(t)=\{u_k(t), \ k>k_2^{'}\}, \ \forall \ t= 1,\ldots,N$ denote the rejected noise BLIMFs to differentiate these noise modes from the relevant signal modes. First, we divide each rejected noise mode $u_k^{(n)}(t)$ into $M$ non-overlapping segments/ vectors $\pmb{u}_{jk}^{(n)}=\{u_k^{(n)}(t)\ \forall \ t=j-\frac{L}{2},\ldots,j+\frac{L}{2}\}$ of equal size $L+1$ and centered around an index $j$ (depicted using boxes drawn on the rejected noise coefficients at the bottom left of Fig. \ref{Fig:05}). Next, the EDF $E_{jk}^{(n)}(z)$ of the noise segment $\pmb{u}_{jk}^{(n)}$ from the $k$th rejected BLIMF $u_k^{(n)}(t)$, is computed as follows
\begin{align}\label{EDFj}
    E_{jk}^{(n)}(z) = \frac{1}{L+1}\sum_{t=1}^{L+1} (u_k^{(n)}(t)\leq z), \ \ k>k_2^{'}.
\end{align} 
This way, EDFs of all of the non-overlapping segments are computed using \eqref{EDFj}, This step is depicted using the $EDF$ blocks in Fig. \ref{Fig:05}.

Finally, by ensemble averaging the EDFs of all the segments from the rejected modes yields a close estimate $\hat{E}_0(z)$ of the actual noise CDF
\begin{align}\label{Eq:18}
    \hat{E}_0(z) = \frac{1}{M(K-k_2^{'})}\sum_{k=k_2^{'}+1}^{K}\ \sum_{j=1}^{M}E_{jk}^{(n)}(z),
\end{align}
where the accuracy of the estimate $\hat{E}_0(z)\approx E_0(z)$ increases with increase in number of segments, i.e., the length of the dataset. 
Fig. \ref{Fig:05} plots the resulting estimate $\hat{E}_0(z)$ of the noise CDF $E_0(z)$ that is obtained from the rejected modes of the noisy signal corrupted by additive wGn at SNR $=10$ dB.

\subsection{Thresholding Relevant Modes Using CVM Test}
This section describes the CVM statistic-based testing framework used to reject noise from the selected relevant modes $\{u_k(t), \ k\leq k_2^{'}\}$. The aim here is to reject the coefficients corresponding to noise $\psi(t)$ without losing those corresponding to the true signal $x(t)$. In this regard, detection of noise coefficients from the selected modes is defined as a local hypothesis testing problem by selecting a local segment $\pmb{u}_{jk}=\{u_{k}(t) \ \forall \ t =j-L/2,\ldots,j+L/2\}$ of size $L+1$, around each coefficient $u_k(j)$, from a selected BLIMF $\{u_k (t), \ k\leq k_2^{'}\}$, as follows
\begin{align}\label{Eq:19} \nonumber
    & \hat{\mathcal{H}}_0 : \pmb{u}_{jk}\in \psi(t), \\
    & \hat{\mathcal{H}}_1 :  \pmb{u}_{jk}\in x(t),
\end{align}
where $\hat{\mathcal{H}}_0$ and $\hat{\mathcal{H}}_1$ respectively denote the null and alternate hypothesis of our VMD-based denoising approach.

In order to test the hypothesis given in \eqref{Eq:19}, i.e, to check the possibility that $\pmb{u}_{jk}\in \psi(t)$; the EDF $E_{jk}(z)$ of the local segment $\pmb{u}_{jk}$ is computed based on \eqref{EDFj} and then the goodness of fit (GoF) of $E_{jk}(z)$ is evaluated/ tested on the estimated noise EDF $\hat{E}_0$, where
\begin{align}\label{Eq:20} 
    &\mathcal{H}_0 : E_{jk}(z)\sim \hat{E}_0(z) \Rightarrow \Delta_{jk} \leq \lambda_k,\nonumber \\
    &\mathcal{H}_1 : E_{jk}(z)\nsim \hat{E}_0(z) \Rightarrow \Delta_{jk} > \lambda_k.
\end{align}
Here, the symbol $\sim$ denotes the \emph{close-fit} and $\nsim$ denotes \emph{no-fit} of EDFs which is decided based on the value of the CVM distance $\Delta_{jk}$ between $E_{jk}(z)$ and $\hat{E}_0(z)$ computed using \eqref{Eq:05}. 
To achieve that, a threshold $\lambda_k$ is estimated such that the false detection of null hypothesis $\mathcal{H}_0$, referred to as false alarms, are minimized. In essence, $\lambda_k$ indicates the maximum possible value of distance $\Delta_{jk}$ required to suggest a \emph{close-fit} between the two EDFs. 
Therefore, a \emph{close-fit} $E_{jk}(z)\sim \hat{E}_0(z)$ is detected when $\Delta_{jk}$ is within the specified bounds, i.e., $\Delta_{jk} \leq \lambda_k$. On the other hand, the case of \emph{no-fit} ${E}_{jk}(z)\nsim \hat{E}_0(z)$ is obtained when the distance exceeds the specified bound of threshold, i.e.,  $\Delta_{jk} > \lambda_k$. 

\begin{figure*}[t]
	\begin{minipage}[b]{0.3\linewidth} \centerline{\includegraphics[scale=0.265]{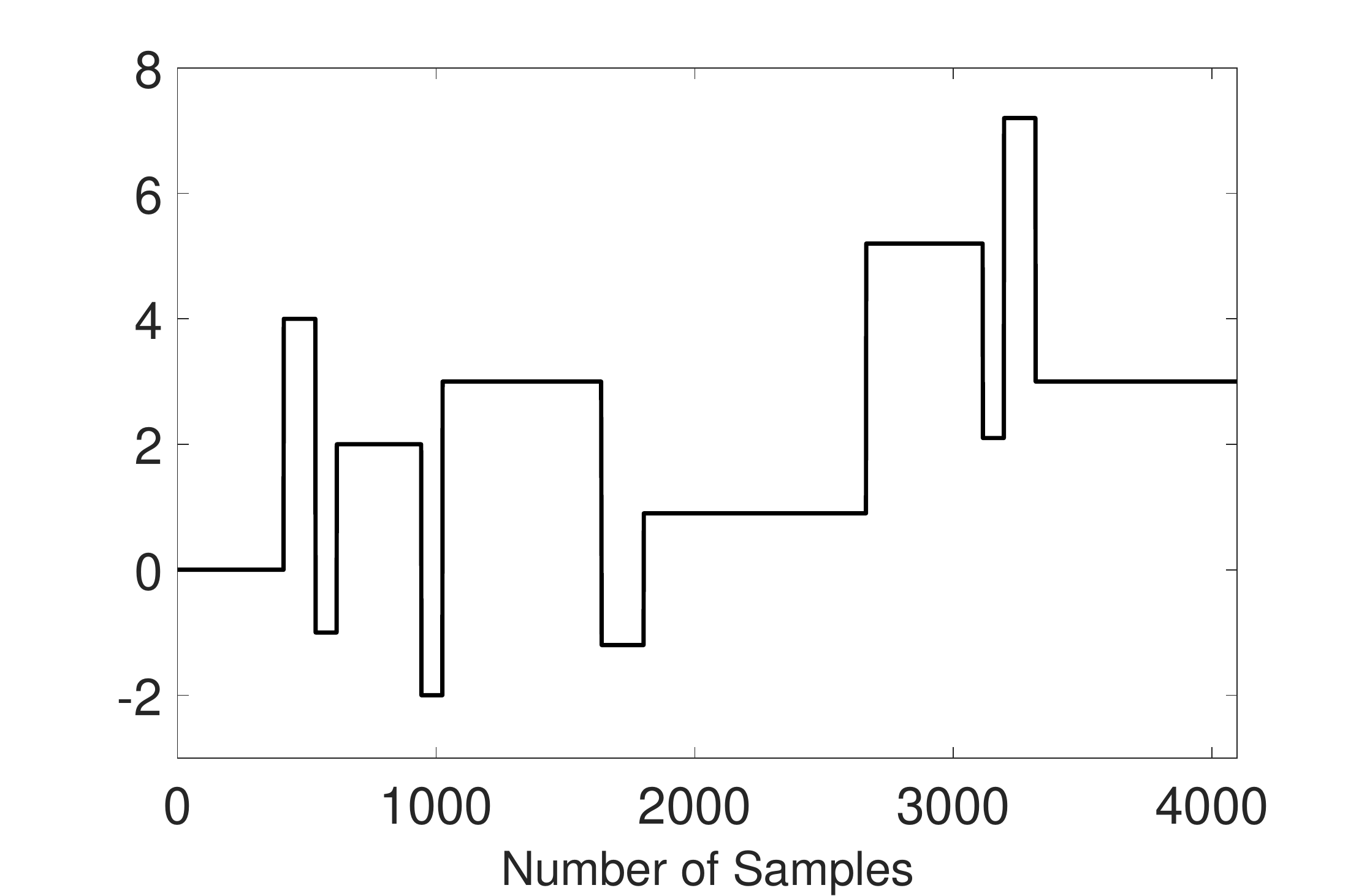}} \centerline{(a) \small{Blocks}} \end{minipage}
	\begin{minipage}[b]{0.3\linewidth} \centerline{\includegraphics[scale=0.265]{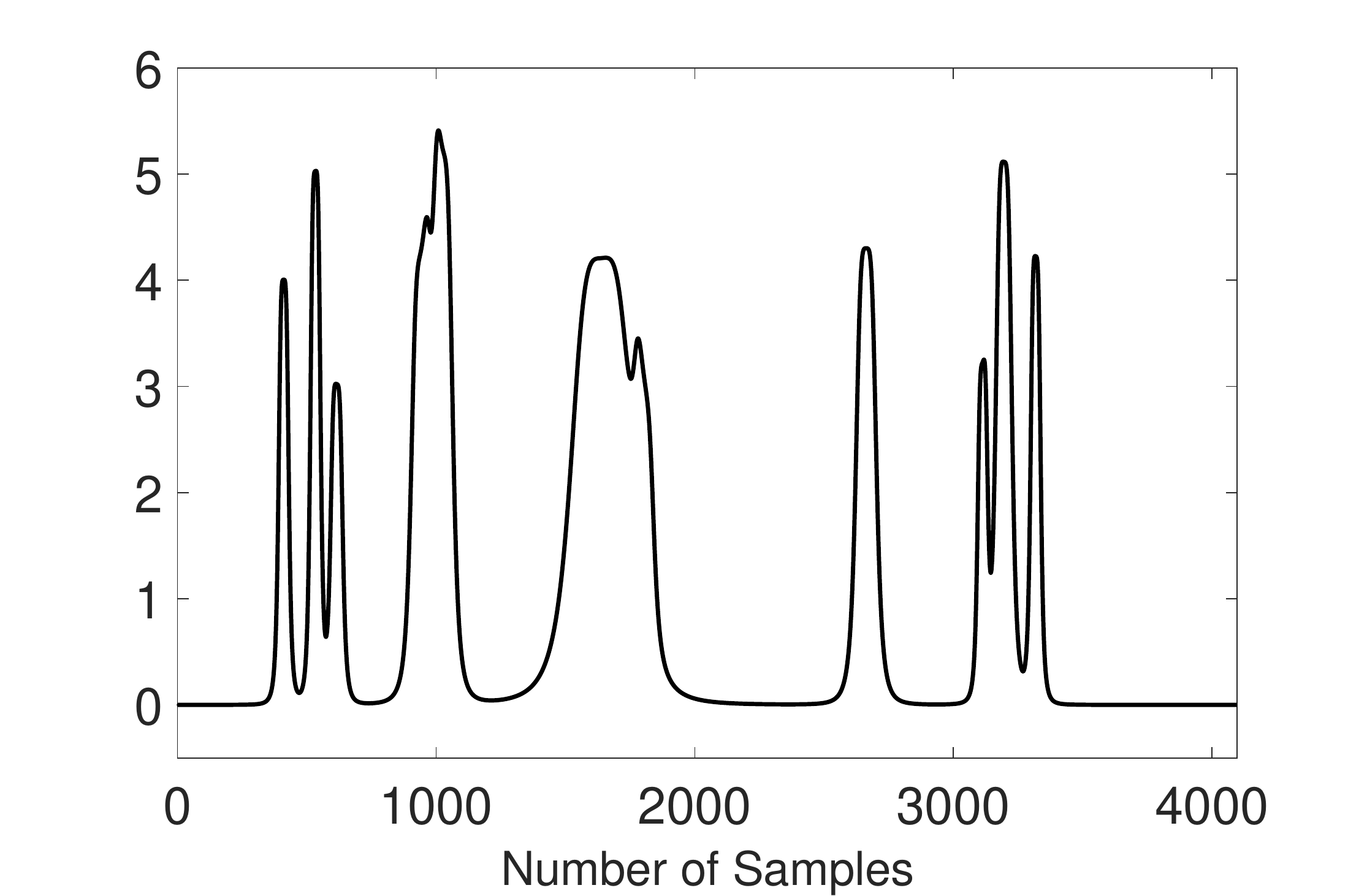}} \centerline{(b) \small{Bumps}} \end{minipage}
	\begin{minipage}[b]{0.3\linewidth} \centerline{\includegraphics[scale=0.265]{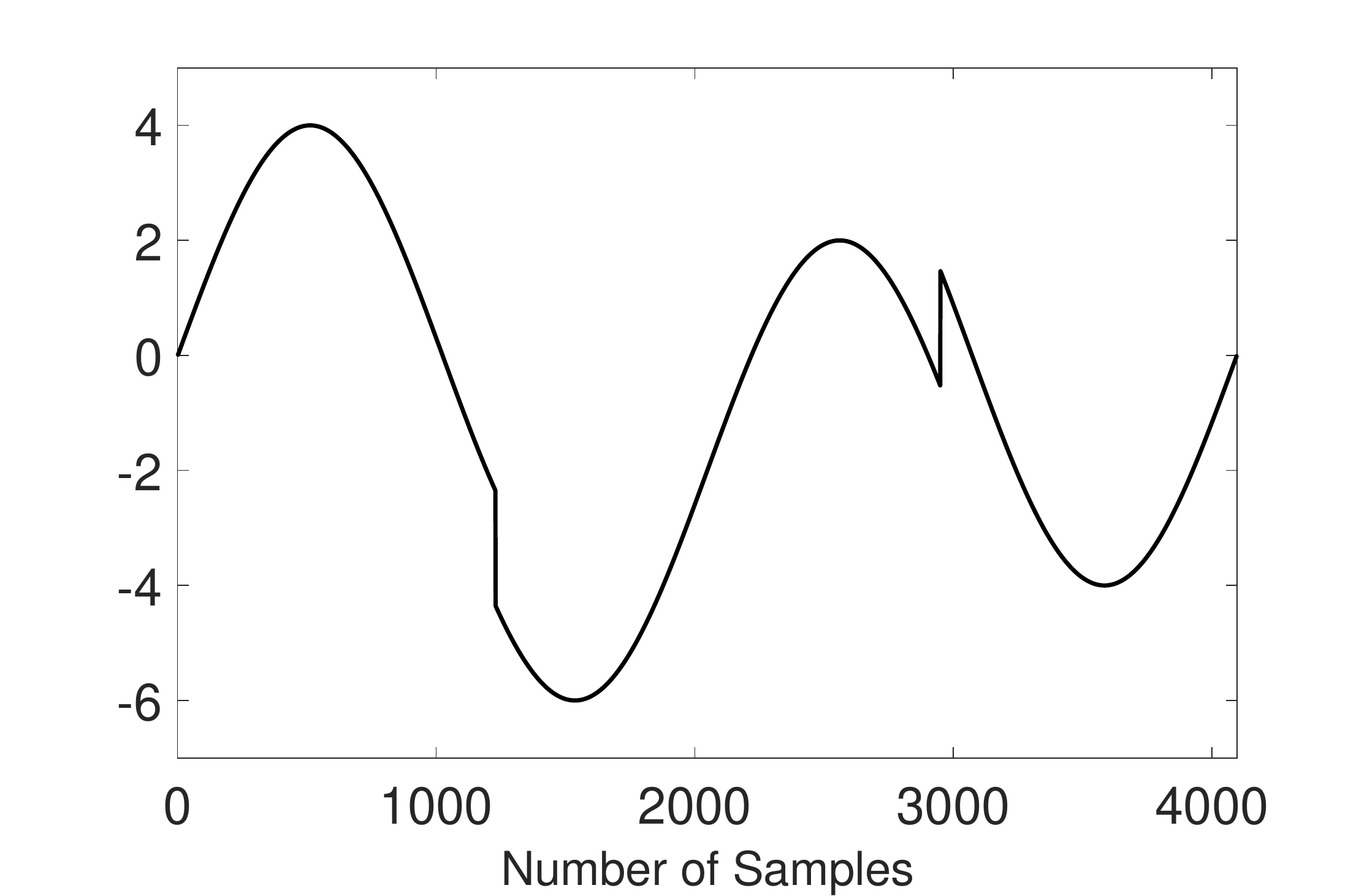}} \centerline{(c) \small{Heavy Sine}} \end{minipage}\\
	\begin{minipage}[b]{0.3\linewidth} \centerline{\includegraphics[scale=0.265]{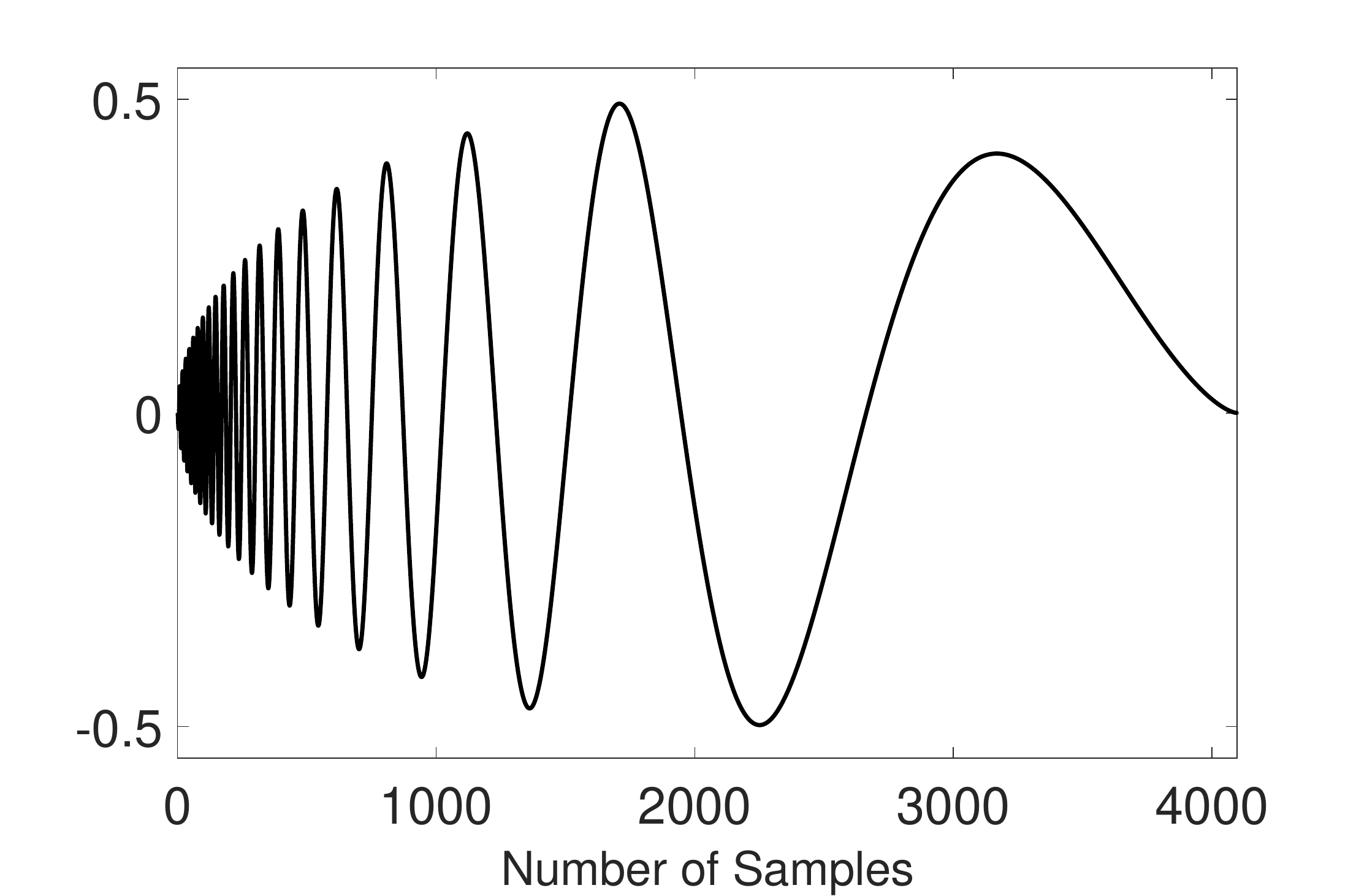}} \centerline{(d) \small{Doppler}} 
	\end{minipage}
	\begin{minipage}[b]{0.3\linewidth} \centerline{\includegraphics[scale=0.265]{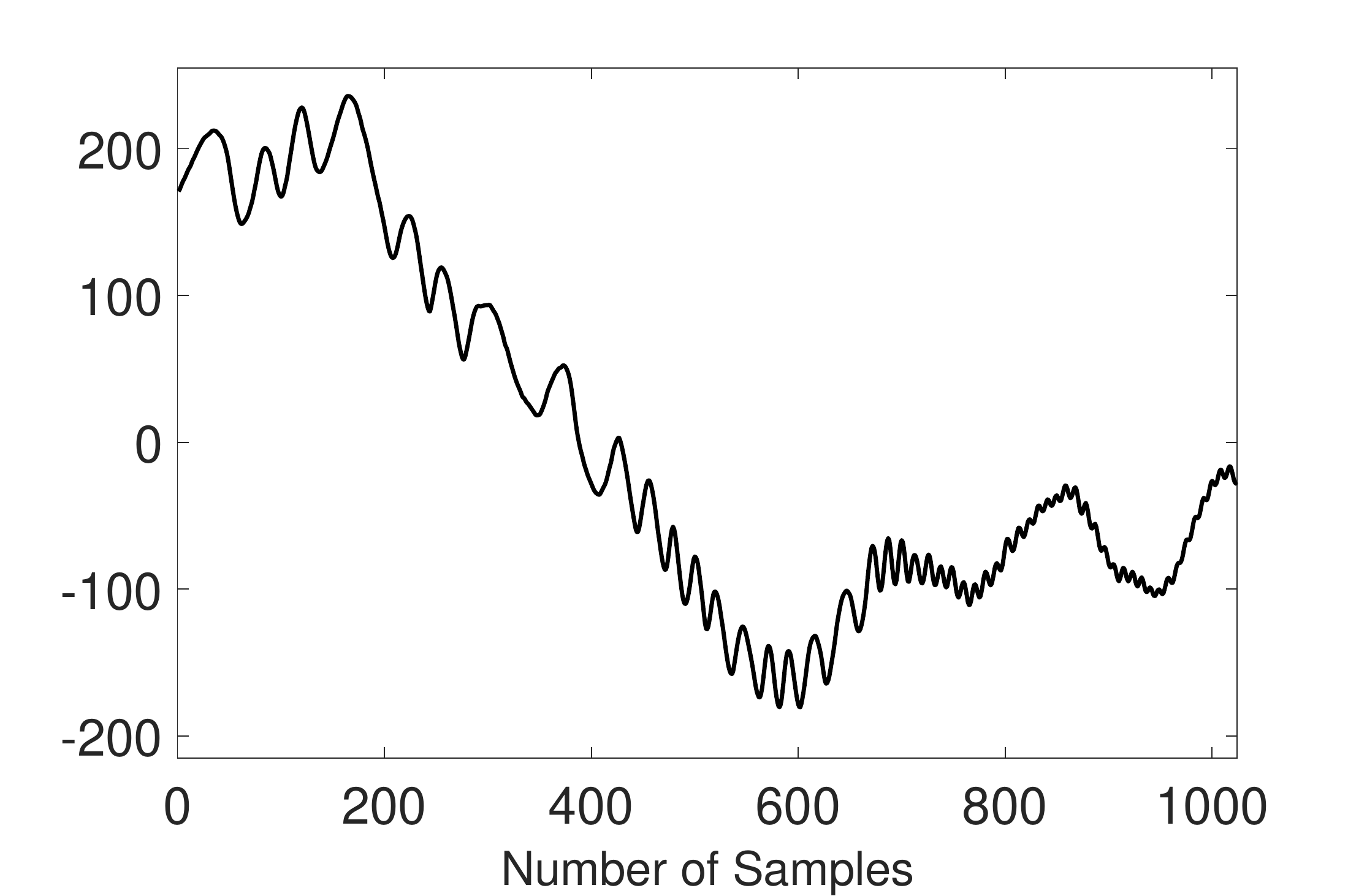}} \centerline{(e) Sofar} \end{minipage}
	\begin{minipage}[b]{0.3\linewidth} \centerline{\includegraphics[scale=0.265]{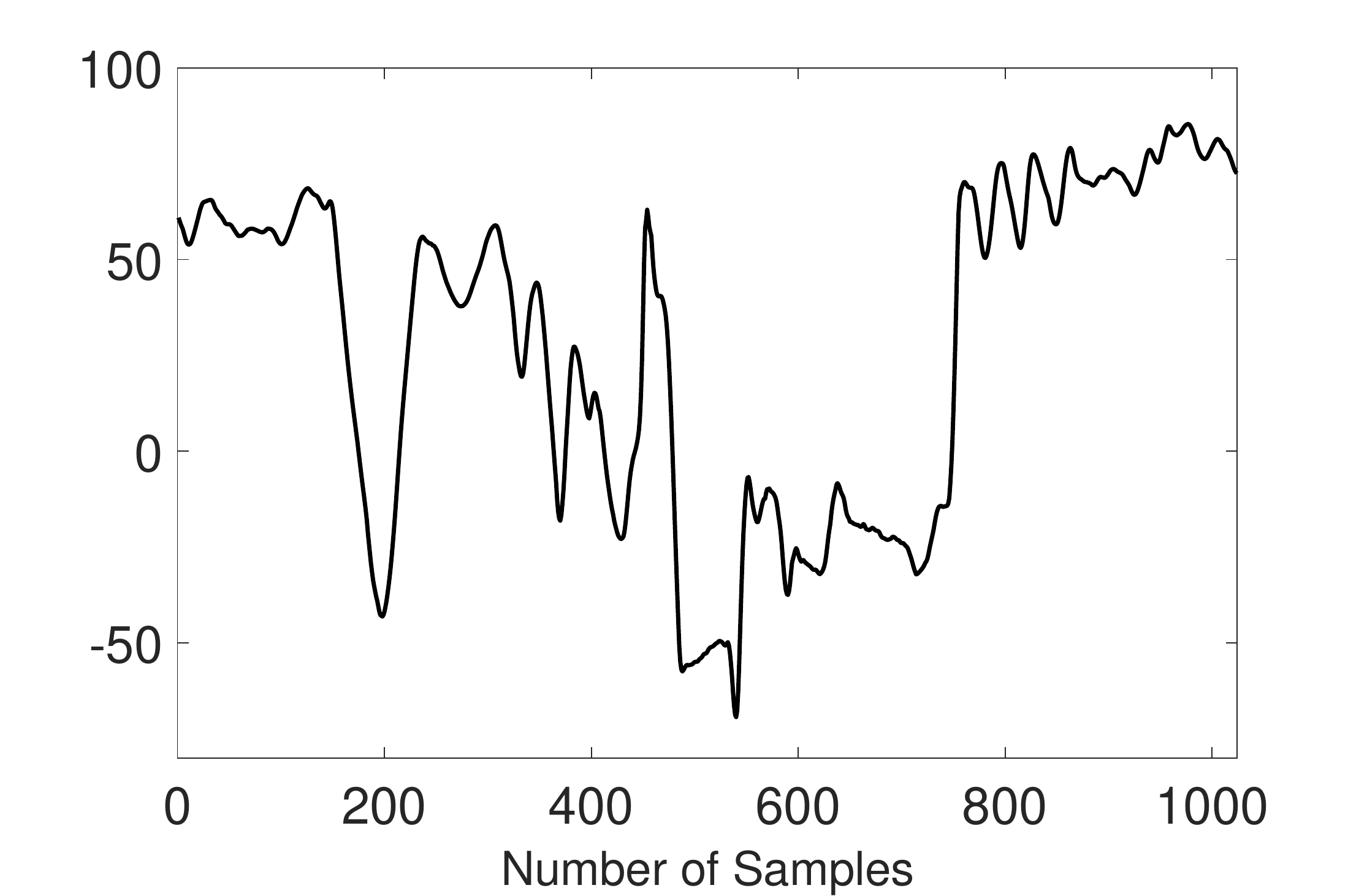}} \centerline{(f) \small{Tai Chi}} \end{minipage}
	\caption{Test signals used for experimentation in this work.}
	\label{Ch1:InputRealSigs}
\end{figure*}
\subsubsection{Selection of Threshold Based on Rejected Noise Modes}
Within GoF tests, the threshold or critical value $\lambda_k$ is selected for each mode $k$ that serves as an upper bound on the CVM distances $\Delta_{jk}$. Generally, $\lambda$ is selected for very small value of $P_{\textrm{fa}}$ \eqref{Eq:07} which ensures least false rejections of the null hypothesis $\mathcal{H}_0$. That means false detections of {\emph{no-fit}} (i.e., $\mathcal{H}_1$) when the reference EDF actually fits the given data samples (i.e., $\mathcal{H}_0$) are minimized. In the context of the testing problem \eqref{Eq:21}, very small $P_{\textrm{fa}}$ means threshold is selected such that it minimizes the false detection of noise (i.e. $\hat{\mathcal{H}}_0$) as true signal (i.e. $\hat{\mathcal{H}}_1$). Hence, this requirement of minimum $P_{\textrm{fa}}$ fits right in the denoising problem since the goal in denoising is to ensure maximum noise is removed which can be achieved by minimizing the $P_{\textrm{fa}}$ while attempting to maximize the preservation of the true signal.

Conventionally, threshold selection is performed by accumulating the probabilities of false detection, i.e., $P_{\textrm{fa}} = \textrm{Prob}(\Delta_{jk}>\lambda | \hat{\mathcal{H}}_0)$. Given the PDF $p(u_{k}^{(n)}(t))$ of rejected noise modes $u_{k}^{(n)}(t)$, $P_{\textrm{fa}}$ is defined as follows
\begin{align}\label{Eq:21}\nonumber
    P_{\textrm{fa}} &  = \textrm{Prob}(\hat{\mathcal{H}}_1 | \hat{\mathcal{H}}_0) \\
                    & = \ \int_{\{u_{k}^\psi(t); \ \Delta_{jk}>\lambda |\hat{\mathcal{H}}_0\}} p(u_{k}^{(n)}(t))du_{jk}^\psi(t),
\end{align}
where $\{u_{k}^{(n)}(t); \ \Delta_{jk}>\lambda |\hat{\mathcal{H}}_0\}$ are noise coefficients from the rejected BLIMFs $\{u_{k}(t), \ k>k_2^{'}\}$ yielding false alarms, i.e., falsely suggesting signal detection. 

An estimate $\hat{p}(u_k^{(n)}(t))$ of the noise PDF $p(u_{k}^{(n)}(t))$ may be obtained by computing the derivative of the empirically estimated noise CDF $\hat{E}_0(t)$ \eqref{Eq:18}
\begin{align}\label{Eq:22}
    \hat{p}(u_k^{(n)}(t)) = \frac{d\hat{E}_0(z)}{dz}.
\end{align}
Consequently, an empirical adaptation of \eqref{Eq:21} to our denoising problem may be obtained through
\begin{align}\label{Eq:23}
    P_{\textrm{fa}} = \ \int_{\{u_{k}^{(n)}(t); \ \Delta_{jk}>\lambda |\hat{\mathcal{H}}_0\}} \frac{d\hat{E}_0(z)}{dt} du_k^{\psi}(t).
\end{align}

Since, the detection of the range of coefficients $\{u_{k}^{(n)}(t); \ \Delta_{jk}>\lambda |\hat{\mathcal{H}}_0\}$ is central to the computation of $P_{\textrm{fa}}$ using \eqref{Eq:24} for a given threshold $\lambda_k$. The proposed empirical approach estimates the coefficients $\{ u_{k}^{(n)}(t); \ \Delta_{jk}>\lambda\}$ from the rejected noise modes $\{u_k(t), \ k> k_2^{'}\}\in \psi(t)$.

To begin with, a range of candidate thresholds are selected and the noise coefficients within each rejected mode $u_k^{(n)}(t)$ are divided into $M$ windows having size $L+1$. Next, each candidate threshold $\lambda$ is used within the CVM test for applying it on the coefficients from each window. Therein, CVM statistic $\Delta_{jk}$ between the EDF $E(z)$ of a window of noise coefficients and the reference noise EDF $\hat{E}_0(z)$ is computed through \eqref{Eq:05} followed by hypothesis testing based on \eqref{Eq:20}. For each $\lambda$, probability of false alarm ${P_{\textrm{fa}}}$ is computed by recording the instances of erroneously detecting noise segments (from rejected modes) as signal and then dividing the accumulated false alarms by total number of windows $M$. 

This way, a threshold versus $P_{\textrm{fa}}$ table is estimated by computing the ${P_{\textrm{fa}}}$ for all the candidate thresholds $\lambda$. The relationship between the threshold $\lambda$ and the ${P_{\textrm{fa}}}$, obtained from the rejected BLIMFs of a given input noisy signal having SNR $=10$ dB, is graphically shown in Fig. \ref{Fig:05} (bottom right). In general, higher number of false alarms (i.e., higher ${P_{\textrm{fa}}}$) are observed for lower thresholds but with increase in threshold value, the ${P_{\textrm{fa}}}$ decreases. 

Afterwards, a threshold is selected for a given ${P_{\textrm{fa}}}$ from the estimated threshold $\lambda$ versus ${P_{\textrm{fa}}}$ table. 
Here, it is important to consider the trade-off between the $P_{\textrm{fa}}$ and the probability of true-signal detection $P_{\textrm{d}}$ when selecting a threshold for noise reduction. Generally, $P_{\textrm{d}}$ also decreases with decrease in $P_{\textrm{fa}}$, i.e., signal is falsely (or erroneously) rejected as noise for lower $P_{\textrm{fa}}$, see \cite{steven993BookDT} for more insight into this matter. Therefore, to avoid loss of signal from the initial BLIMFs (which are mostly composed of signal), higher ${P_{\textrm{fa}}}$ is selected to keep $P_{\textrm{d}}$ high as well, i.e., signal is not falsely rejected as noise. On the other hand, lower $P_{\textrm{fa}}$ is chosen for latter modes (with dominant noise component) to reject maximum noise. To that end, a scale adaptive separate threshold $\lambda_k$ is selected for each of the relevant mode $u_k(t)$ using the following decaying function
\begin{align}\label{Eq:24}
    P_{\textrm{fa}}^{(k)} = e^{-k+1},
\end{align}
where $P_{\textrm{fa}}^{(k)}$ denotes the false alarm probability of the $k$th mode. The decaying function \eqref{Eq:24} assigns higher $P_{\textrm{fa}}$ to initial signal modes (i.e., to recover maximum signal) and smaller $P_{\textrm{fa}}$ to latter noise BLIMFs (i.e., to reject maximum noise).

\begin{table*}[t]
	\caption{Output SNR values obtained from various denoising methods for varying input SNR levels.}
	\centering
	\resizebox{\textwidth}{!}{
		\begin{tabular}{ll|llll|llll|llll|llll}\thickhline
			
			\textbf{Input SNR} & &\textbf{-5} & \textbf{0} & \textbf{5} & \textbf{10} & \textbf{-5} & \textbf{0} & \textbf{5} & \textbf{10} & \textbf{-5} & \textbf{0} & \textbf{5} & \textbf{10} & \textbf{-5} & \textbf{0} & \textbf{5} & \textbf{10}\\ \thickhline
			&  & &  &  &  &  &  &  &  &  &  &  &  &  &  &  &  \\
			\textbf{Inp. Signal}&&\multicolumn{3}{c}{\textbf{Blocks}} &&\multicolumn{3}{c}{\textbf{Bumps}} &&\multicolumn{3}{c}{\textbf{Sofar}} &&\multicolumn{3}{c}{\textbf{Tai Chi}} &\\
			&  & &  &  &  &  &  &  &  &  &  &  &  &  &  &  &  \\
			\textbf{EMD-DFA} & SNR & -2.00 & 2.83 & 6.39 & 7.52 & 2.78 & 7.77 & 12.85 & 17.11 & -1.43 & 3.24 & 8.41 & 13.88 & -1.38 & 2.97 & 8.79 & 13.46 \\
			& MSE & 13.96 & 4.64 & 2.01 & 1.55 & 2.78 & 7.77 & 12.85 & 17.11 & 2.01e4 & 0.68e4 & 0.21e4 & 0.05e4 & 0.36e4 & 0.13e4 & 0.03e4 & 0.01e4 \\
			& & &  &  &  &  &  &  &  &  &  &  &  &  &  &  &  \\
			\textbf{VMD-DFA} & SNR & 2.77 & 7.67 & 12.71 & 17.31 & 2.95 & 8.11 & 12.74 & 18.56 & 2.86 & 7.66 & 12.72 & 17.89 & 2.80 & 7.43 & 12.98 & 18.19 \\
			& MSE & 4.65 & 1.50 & 0.47 & 0.16 & 1.65 & 0.50 & 0.17 & 0.04 & 0.75e4 & 0.24e4 & 0.07e4 & 0.02e4 & 1.37e3 & 0.47e3 & 0.13e3 & 0.04e3 \\
			&  &  &  &  &  & & &  &  &  &  &  &  &  &  &  &  \\
			\textbf{EMD-IT} & SNR & 5.38 & 9.90 & 14.57 & 18.88 & 6.06 & 10.29 & 15.19 & 19.83 & 4.53 & 9.31 & 13.56 & 17.43 & 3.73 & 8.53 & 12.84 & 17.29 \\
			& MSE &2.56 & 0.90 & 0.30 & 0.11 & 0.80 & 0.30 & 0.09 & 0.03 & 0.52e4 & 0.17e4 & 0.06e4 & 0.02e4 & 1.15e3 & 0.36e3 & 0.13e3 & 0.04e3 \\
			& & &  &  &  &  &  &  &  &  &  &  &  &  &  &  &  \\
			\textbf{DTCWT-Thr} & SNR & 3.37 & 10.71 & 15.36 & 19.01 & 6.14 & 10.95 & 16.12 & 21.05 & -5.05 & 0.11 & 4.92 & 10.09 & -4.97 & 0.06 & 5.13 & 10.75 \\
			& MSE & 4.05 & 0.74 & 0.25 & 0.11 & 0.78 & 0.26 & 0.07 & 0.02 & 0.46e5 & 0.14e5 & 0.04e5 & 0.01e5 & 0.82e4 & 0.25e4 & 0.08e4 & 0.02e4 \\
			& & &  &  &  &  &  &  &  &  &  &  &  &  &  &  &  \\
			\textbf{BLFDR} & SNR & -0.62 & 5.33 & 12.09 & 17.88 & -0.19 & 6.26 & 12.90 & 19.58 & 1.59 & 8.26 & 13.81 & 18.16 & 1.16 & 7.35 & 12.99 & 18.17 \\
			& MSE & 10.18 & 2.58 & 0.54 & 0.14 & 3.39 & 0.76 & 0.16 & 0.03 & 1.01e4 & 0.41e4 & 0.06e4 & 0.02e4 & 2.00e3 & 0.48e3 & 0.13e3 & 0.03e3 \\
			& & &  &  &  &  &  &  &  &  &  &  &  &  &  &  &  \\
			\textbf{DWT-GoF} & SNR & 5.08 & 9.37 & 13.92 & 17.83 & 7.02 & 11.50 & 16.40 & 20.98 & 4.56 & 9.66 & 14.21 & 18.95 & 4.20 & 8.88 & 12.80 & 15.37 \\
			& MSE & 2.73 & 1.01 & 0.35 & 0.14 & 0.64 & 0.23 & 0.07 & 0.02 & 0.50e4 & 0.15e4 & 0.05e4 & 0.01e4 & 1.00e3 & 0.33e3 & 0.13e3 & 0.07e3 \\
			& & &  &  &  &  &  &  &  &  &  &  &  &  &  &  &  \\
			\textbf{DT-GOF-NeighFilt} & SNR & 11.20 & \textbf{15.30} & 17.10 & 19.27 & 10.73 & 15.16 & \textbf{18.52} & \textbf{23.28} & \textbf{10.91} & 14.25 & 16.59 & 17.99 & 9.78 & 12.42 & 15.41 & 18.29 \\
			& MSE & 0.66 & \textbf{0.26} & 0.13 & 0.08 & 0.25 & 0.09 & \textbf{0.04} & \textbf{0.01} & \textbf{1.18e3} & 0.54e3 & 0.31e3 & 0.22e3 & 0.27e3 & 0.15e3 & 0.07e3 & 0.03e3 \\
			& & &  &  &  &  &  &  &  &  &  &  &  &  &  &  &  \\
			\textbf{Prop. VMD-CVM} & SNR & \textbf{11.30} & 15.22 & \textbf{17.21} & \textbf{19.86} &  \textbf{10.92} & \textbf{15.38} & 18.49 & 23.19 & 10.28 & \textbf{14.67} & \textbf{17.35} & \textbf{20.19} & 10.04 & \textbf{12.58} & \textbf{16.19} & \textbf{19.38} \\
			& MSE &\textbf{0.65} & 0.26 & \textbf{0.13} & \textbf{0.07} & \textbf{0.25} & \textbf{0.08} & 0.04 & 0.01 & 1.35e3  & \textbf{0.49e3} & \textbf{0.26e3} & \textbf{0.14e3} & 0.25e3 & \textbf{0.14e3} & \textbf{0.06e3} & \textbf{0.03e3} \\
			\thickhline
		\end{tabular}{}}
	\label{table1}
\end{table*}
\subsubsection{Thresholding Function}
Traditional hard-thresholding function detects noise coefficients based on their smaller amplitudes via a threshold that is well adapted to wavelet denoising. However, the use of a thresholding function that detects noise based on amplitude difference among coefficients is not properly motivated for VMD denoising. This is because VMD denoising methods do not exploit sparsity of multiscale decomposition, instead it estimates the local trend to detect the signal content from noise \cite{kopsinis2009EMD-IT,liu2016VMDDFA}. The proposed approach works on the same principle of estimating local trend in a BLIMF to detect and reject noise. In this work, the local trend is estimated using the EDF ${E}_{tk}(z)$ of local segment $\pmb{u}_{tk}$ that is selected around each coefficient at location $t$. Subsequently, it is checked whether the ${E}_{tk}(z)$ is close to that of noise $\hat{E}_0(z)$ by estimating the (CVM) distance $\Delta_{tk}$ between the two EDFs using \eqref{CVMy} and then it is compared against the threshold $\lambda_k$
\begin{align}
	\hat{u}_k(t) =
	\begin{cases}
		0        & \quad    \textrm{if} \ \ \Delta_{tk} \leq \lambda_k,\\
		u_{k}(t), & \quad    \textrm{if} \ \ \Delta_{tk} > \lambda_k.
	\end{cases}
	\label{thr1}
\end{align}
Finally, the denoised signal is reconstructed from the thresholded relevant BLIMFs $\{\hat{u}_{k}, \ k<k_2^{'}\}$, as follows
\begin{align}\label{Eq:26}
	\hat{x}(t)=\sum_{k=1}^{k_2^{'}} \hat{u}_{k}(t), \ \ k<k_2^{'},
\end{align}
where $\hat{x}$ is an estimate of the true signal $x(t)$ obtained using the proposed approach. In the rest of the paper, we will refer to the proposed approach as \textit{VMD-CVM}.
\section{Simulation Results and Discussion}
\label{results}
In this section, simulation results are presented to demonstrate the effectiveness of the performance of the proposed method. Following methods have been considered for the performance comparison.
\begin{itemize}
  \item DTCWT-Thr \cite{selesnick2005DTCWT}: Exploits the quasi-translation invariance of the DTCWT using a nonlinear thresholding function.
  \item BLFDR \cite{lavrik2008BLFDR}: Performs noise shrinkage by employing the false discovery rate (FDR) \cite{abramovich1995Hypotest_FDR1} within the Bayesian framework.
  \item EMD-IT \cite{kopsinis2009EMD-IT}: Performs interval thresholding on the EMD modes of noisy signal to reject noise.
  \item EMD-DFA \cite{mert2014DFA}: Performs partial reconstruction by rejecting the EMD modes exhibiting detrend (i.e., randomness) using the DFA. 
  \item VMD-DFA \cite{liu2016VMDDFA}: Performs partial reconstruction by rejecting the VMD-BLIMFs exhibiting detrend (i.e., randomness) using the DFA.
  \item DWT-GOF \cite{ur2017DWTGoF}: Tests the normality DWT coefficients to detect and reject noise. 
  \item DT-GoF-NeighFilt \cite{naveed2018DTCWTGoF}: Exploits the quasi-translation invariance of the DTCWT using the normality test and neighborhood classification based filtering for effective noise removal. 
\end{itemize}
The following performance measures have been employed for the performance comparison:
\begin{itemize}
  \item Signal-to-noise ratio (SNR);
  \item Mean squared error (MSE).
\end{itemize} 
The test datasets include both real and synthetic signals. Among those, synthetic signals include `Blocks', `Bumps', `Heavy Sine', and `Doppler' respectively plotted in Fig. \ref{Ch1:InputRealSigs} (a)-(d). The real signals include `Sofar' and `Tai Chi' as shown in Fig. \ref{Ch1:InputRealSigs} (e) and (f), respectively whereby prior signal records the oceanographic float drift of the water flowing through Mediterranean sea \cite{richardson1989estbsn} and latter signal tracks the human body movements in a Tai Chi sequence using a 3D sensor attached to the ankles \cite{rehman2009MEMD}.

\begin{figure*}[t]
	\begin{minipage}[b]{0.49\linewidth} \centerline{\includegraphics[scale=0.35]{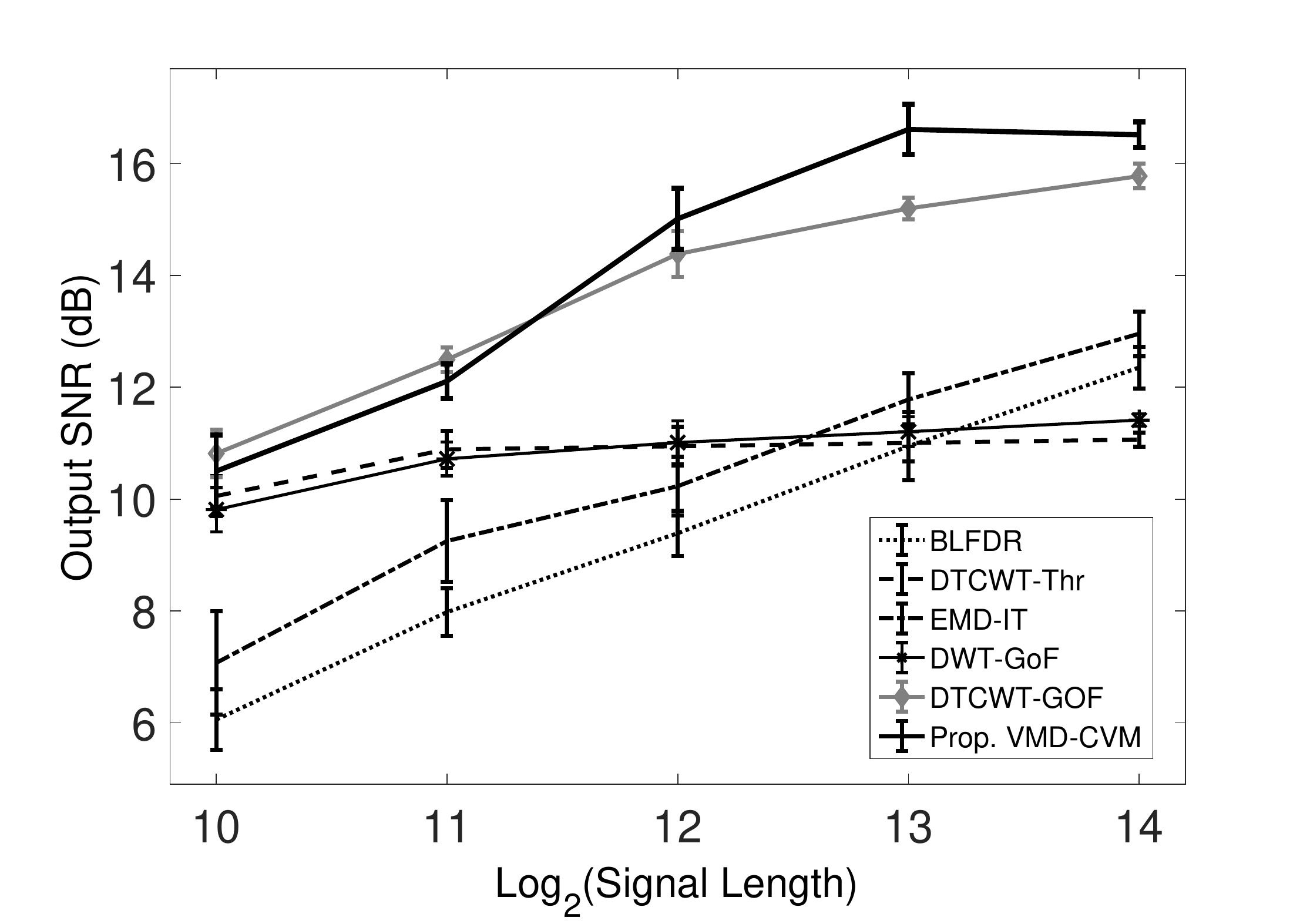}} \centerline{(a) \small{Blocks}} \end{minipage}
	\begin{minipage}[b]{0.49\linewidth} \centerline{\includegraphics[scale=0.35]{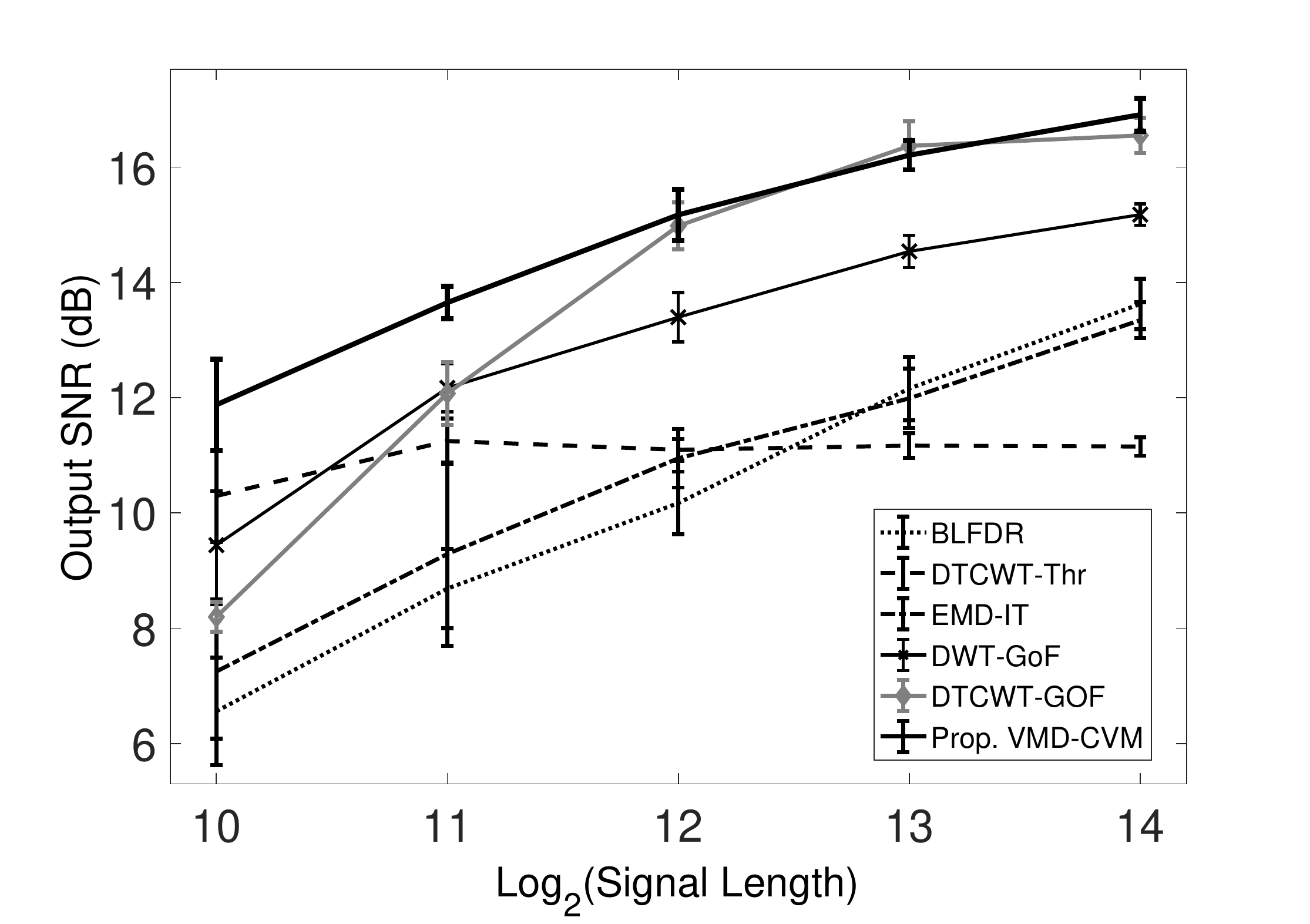}} \centerline{(b) \small{Bumps}} \end{minipage}\\
	\begin{minipage}[b]{0.49\linewidth} \centerline{\includegraphics[scale=0.35]{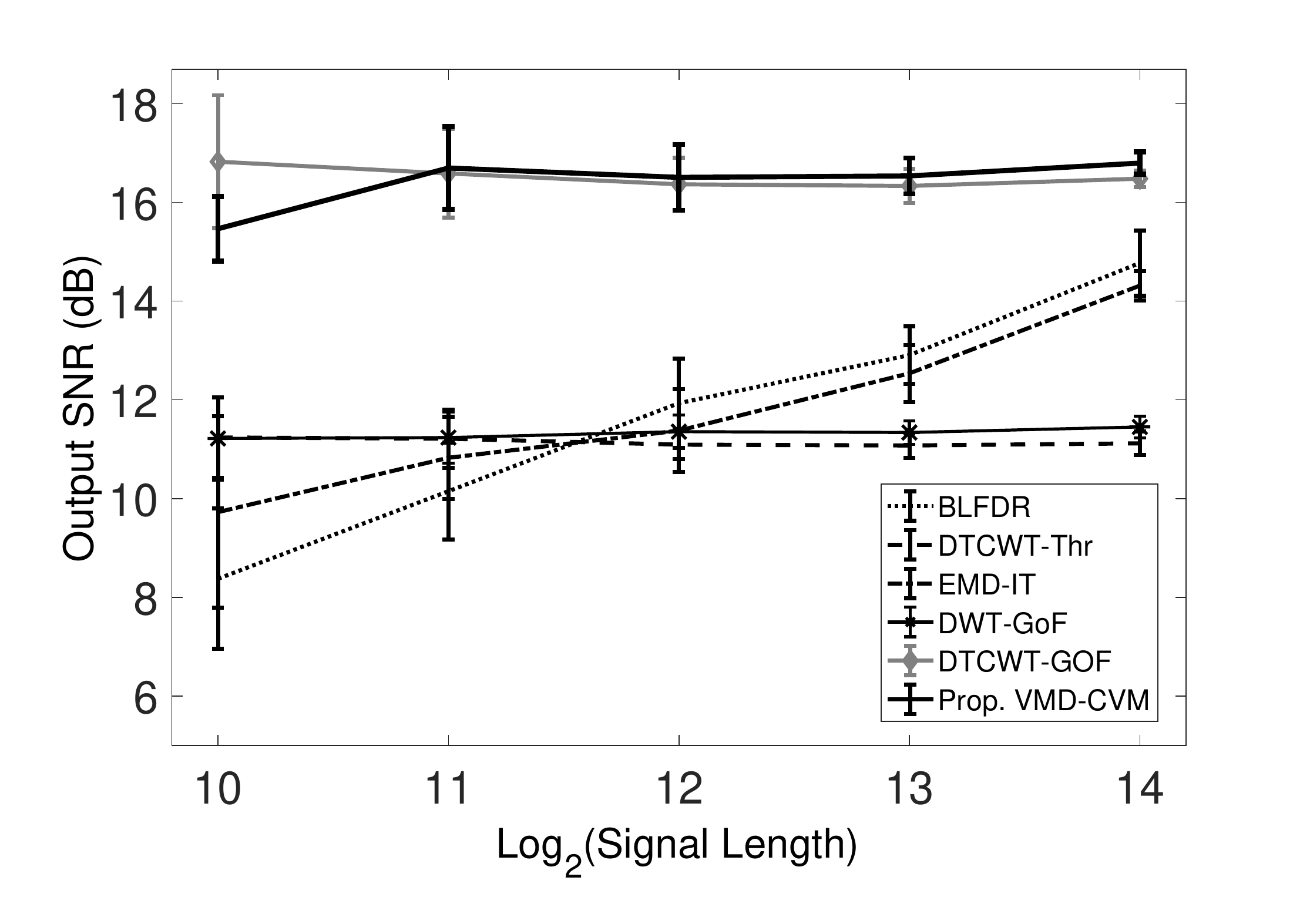}} \centerline{(c) \small{Heavy Sine}} \end{minipage}
	\begin{minipage}[b]{0.49\linewidth} \centerline{\includegraphics[scale=0.35]{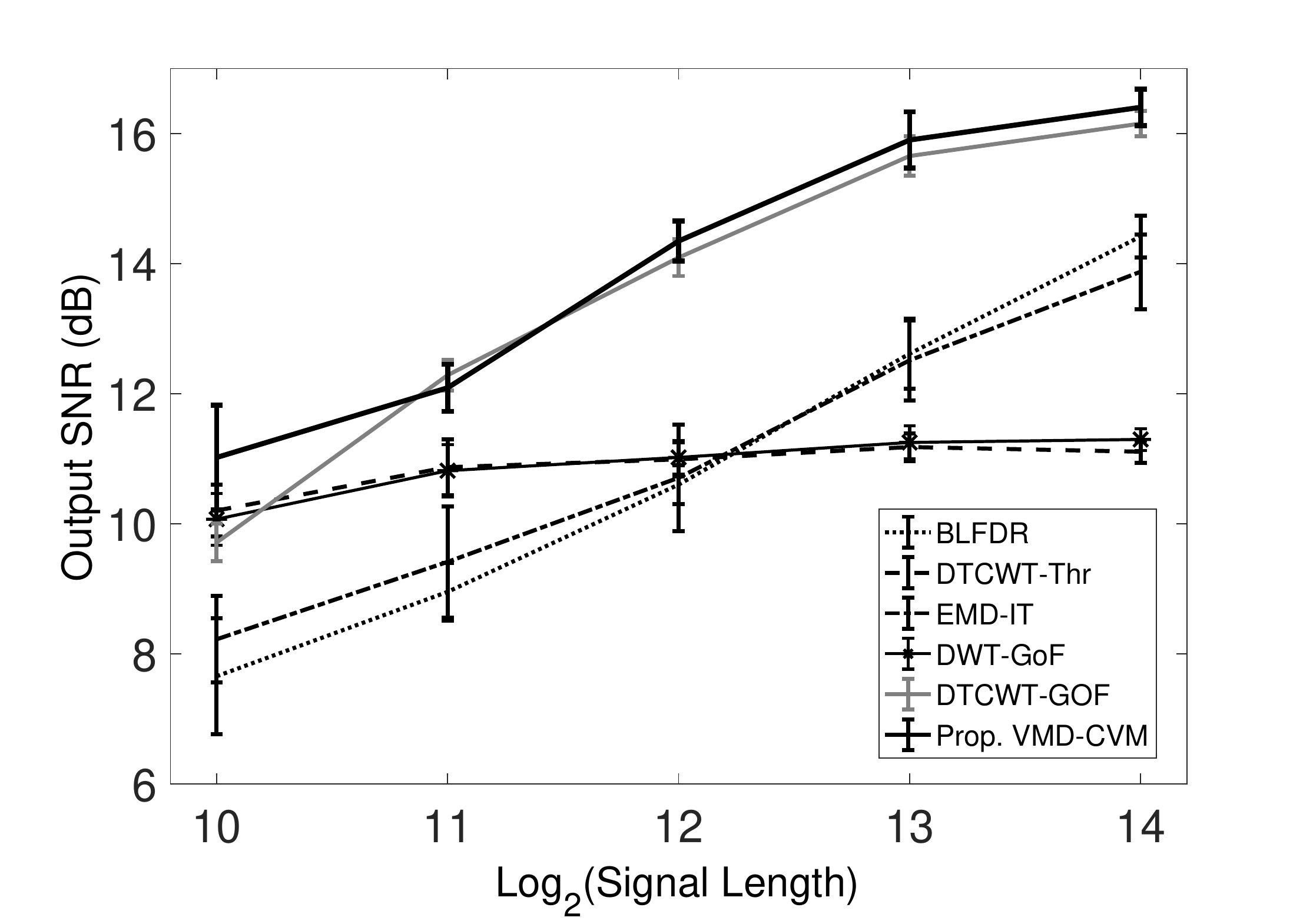}} \centerline{(d) \small{Doppler}} \end{minipage}
	\caption{Error bar plots depicting output SNR levels obtained from different denoising methods for (a) Blocks, (b) Bumps, (c) Heavy sine, (d) Doppler, at input $SNR = 0 dB$ and and varying signal lengths $N=2^{10}\rightarrow2^{14}$}
	\label{Ch1:LenVoSNR}
\end{figure*}
\subsection{Experimental Settings}
We report several experiments to study and analyze different aspects of the proposed method when compared against the sate of the art. In this regard, noisy signals are generated by adding wGn (at varying input SNRs) to the aforementioned input signals shown in Fig. \ref{Ch1:InputRealSigs}. These noisy signals are subsequently denoised using the comparative methods where quantitative measures of performance (i.e., SNR and MSE) are obtained by comparing the clean input signal against the denoised, which are reported in tabular as well as graphical form. The qualitative analysis of the comparative methods is presented by visually demonstrating how closely  the denoised signals follow their corresponding clean input signals.

For VMD-based denoising methods, we chose the number of user defined modes $K=10$ unless specified otherwise by the method. For wavelet denoising, the multiscale decomposition was performed using Daubechies filter bank with eight vanishing moments (i.e., `db8') with number of decomposition levels $M=5$. On the contrary, complex wavelet filters were used when decomposition using the DTCWT. The rest of the simulation parameters for the various methods have been selected on the basis of guidelines provided in the respective references.

\subsection{Input SNR vs. Output SNR}
Table \ref{table1} reports output SNR and MSE values of the denoised signals obtained for various methods considered in this paper. In this regard, input signals including synthetic 'Bumps' and 'Blocks' signals (of $2^{12}$ sample size) and real 'Sofar' and 'Tai Chi' signals (of $2^{10}$ sample size) are corrupted by wGn such that input SNRs become $-5$ dB, $0$ dB, $5$ dB and $10$ dB. For each method, the mentioned output SNR and MSE values in the Table \ref{table1} are average for $J=20$ realizations.  
The best results, i.e., highest output SNR and the corresponding MSE values are highlighted in bold.

It can be seen from the Table. \ref{table1} that the proposed VMD-CVM method outperforms the state of the art methods for almost all input signals, except for a few cases where the DT-GOF-NeighFilt yields better performance than the proposed VMD-CVM. The rest of the data-driven denoising methods based on VMD or EMD fall behind the wavelet denoising methods used in this study. This superior performance of the proposed method owes to the robust multistage procedure that recovers signal within the noisy modes which are rejected as noise in other VMD-based methods.

An important observation in Table. \ref{table1} is the highest output SNR values of the proposed VMD-CVM for the `Blocks' signal despite its piece-wise constant nature. This is significant because VMD/EMD-based denoising methods conventionally fail to extract the details of piece-wise constant signals. This result demonstrates the efficacy of the robust architecture within the proposed approach. For the `Sofar' and `Tai Chi' signals with a fair bit of complexity due to sharp and subtle variations within, the VMD-CVM method yields best performance for input SNR $\geq 0$ dB. Thereby, the margin of difference between the SNR values from proposed method and the DT-GOF-NeighFilt are large enough to be more than $5\%$ of the best.
\begin{figure}[t]
	\begin{minipage}[b]{1\linewidth} \centerline{\includegraphics[scale=0.22]{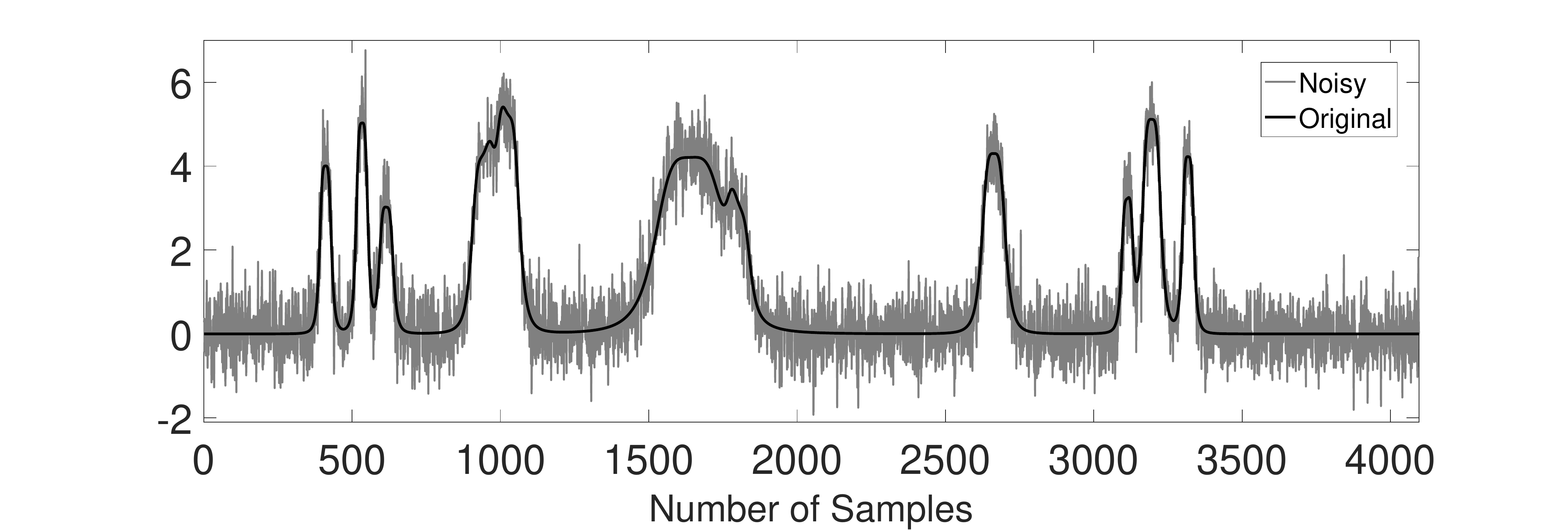}} \centerline{(a) \small{Noisy signal}} 
	\end{minipage}
	\begin{minipage}[b]{1\linewidth} \centerline{\includegraphics[scale=0.22]{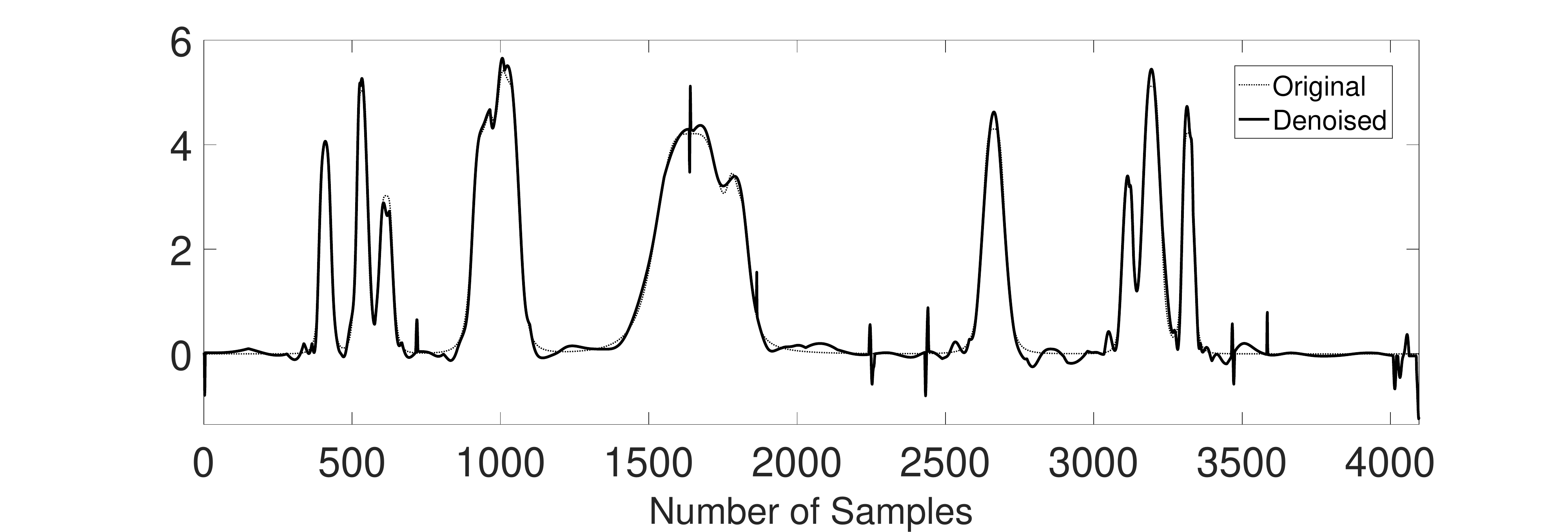}} \centerline{(b) \small{EMD-IT}} \end{minipage}
	\begin{minipage}[b]{1\linewidth} \centerline{\includegraphics[scale=0.22]{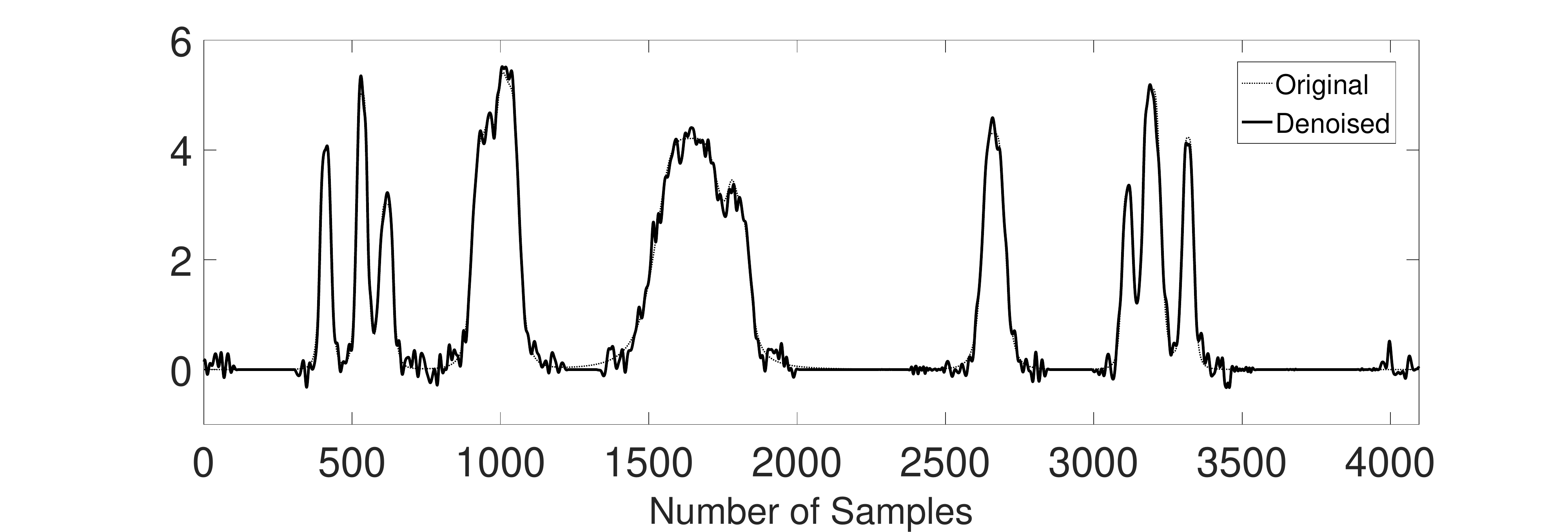}} \centerline{(c) \small{DWT-GoF}} \end{minipage}
	\begin{minipage}[b]{1\linewidth} \centerline{\includegraphics[scale=0.22]{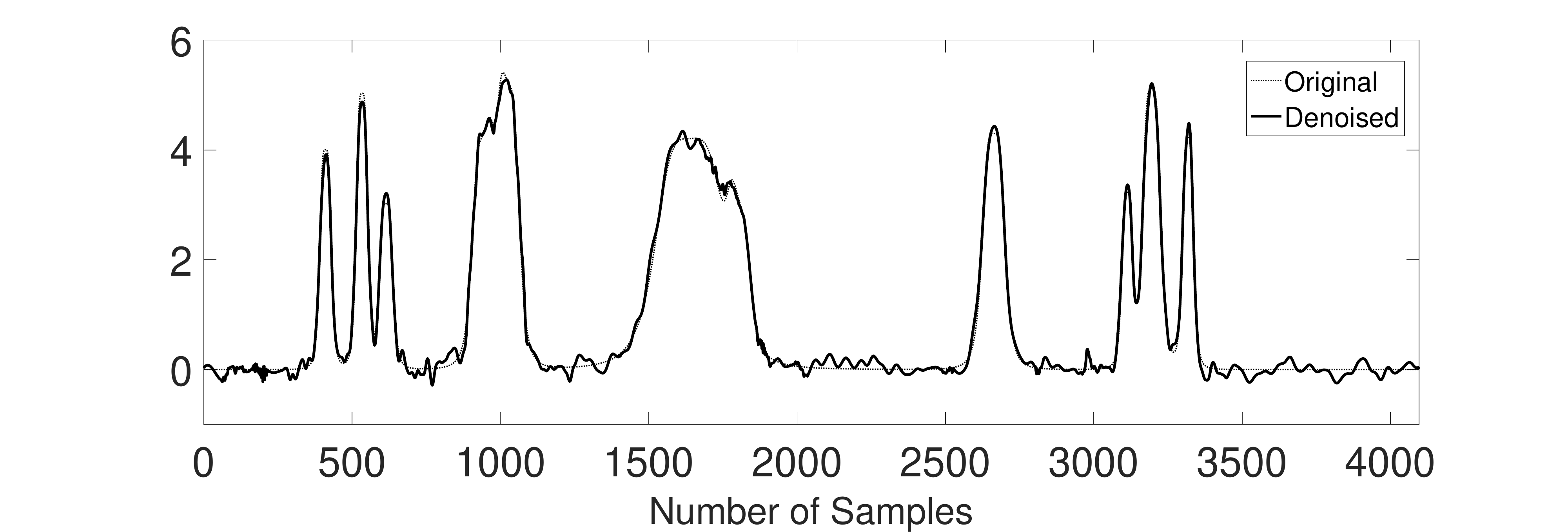}} \centerline{(d) \small{DT-GoF-NeighFilt}} \end{minipage}
	\begin{minipage}[b]{1\linewidth} \centerline{\includegraphics[scale=0.22]{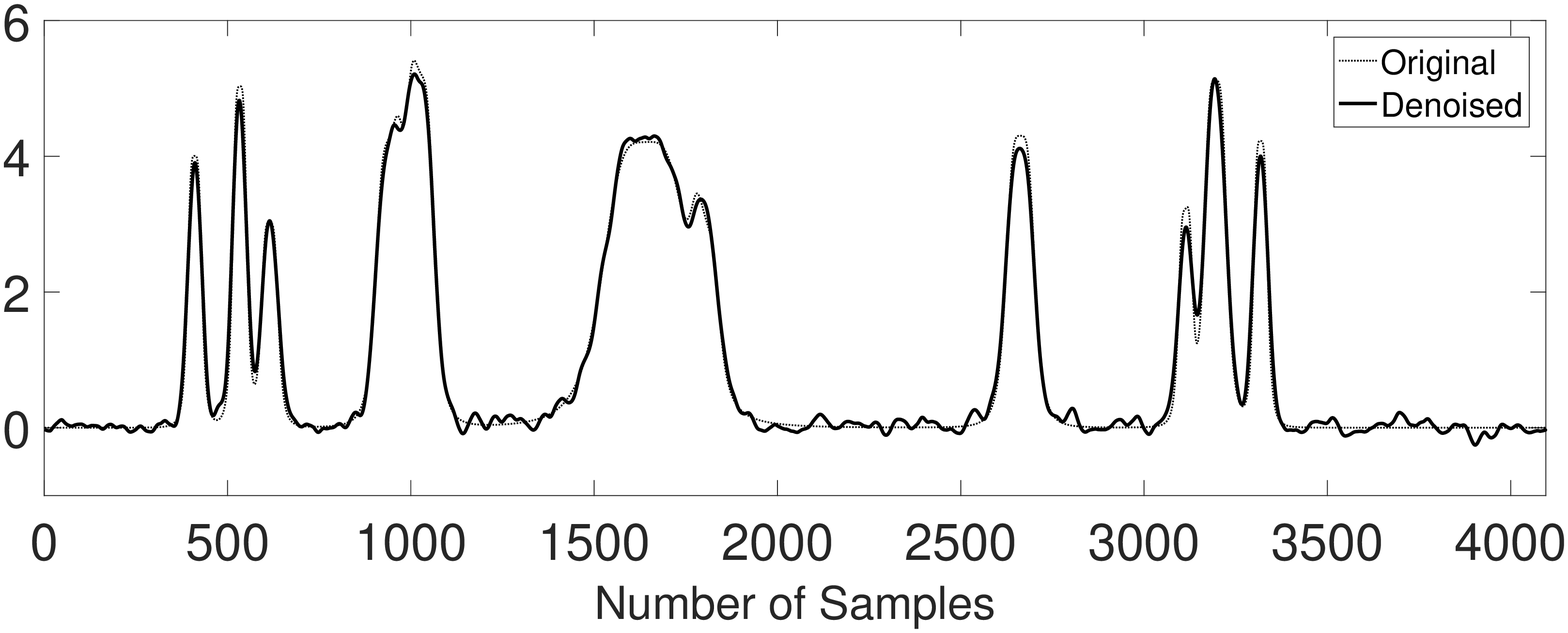}} \centerline{(e) \small{Prop. VMD-CVM}} \end{minipage}
	\caption{Noisy `Bumps' signal (a) and Denoised `Bumps' signals obtained from  various methods ((b) BLFDR, (c) EMD-IT, (d) DWT-GoF, (e) DT-GOF-NeighFilt, and (f) proposed VMD-CVM) for $SNR = 10$ dB and signal lengths $N=2^{12}$.}
	\label{Ch1:DenoisedSet1}
\end{figure}
\subsection{Signal Length vs. Output SNR}
Next, we analyze the proposed framework by comparing it against the state of the art in signal denoising on synthetic signals with varying lengths $N=2^{10}$ to $2^{14}$, corrupted by input noise of $0$ dB. In this experiment, comparative analysis of quantitative results is presented graphically using error-bar plot as shown in Fig. \ref{Ch1:LenVoSNR} that not only displays the mean of the output SNRs over $J=20$ iterations but also gives an estimate of possible variations in SNRs during these iterations. Here, generally it is observed that performance of the denoising methods is significantly improved as the length of the signal increases. That is understandable because increase in length of the signal increases its redundancy that helps in better extraction of signal details in presence of noise.

The error-bar plot for `Blocks' signal in Fig. \ref{Ch1:LenVoSNR} (a) demonstrates that proposed VMD-CVM shows best results for signal length $N\geq 2^{12}$, while DTCWT-GoF yields highest SNRs for $N< 2^{12}$. Similarly, For `Bumps' signal in Fig. \ref{Ch1:LenVoSNR} (b), proposed VMD-CVM yields highest output SNRs for all signal lengths except $N=2^{13}$ where DT-GOF-NeighFilt marginally betters our method. For `Heavy Sine' and `Doppler' signal in Fig. \ref{Ch1:LenVoSNR} (c \& d), the VMD-CVM and DT-GOF-NeighFilt methods closely follow each other and outperform the rest of the comparative methods by a significant margin while yielding similar results on all the lengths. Among these two methods, the proposed method yields highest mean output SNR along with higher standard deviation apart from the odd case where DT-GOF-NeighFilt yields better results.

It can be concluded from the results that proposed VMD-CVM methods stands out along with the DT-GOF-NeighFilt that yielded equally effective denoising performance. Mostly, the VMD-CVM yielded top SNR values especially for higher length input signals. For lower length $N=2^{10}$, DT-GOF-NeighFilt generally outperformed the proposed method.
\begin{figure}[t]
	\begin{minipage}[b]{1\linewidth} \centerline{\includegraphics[scale=0.22]{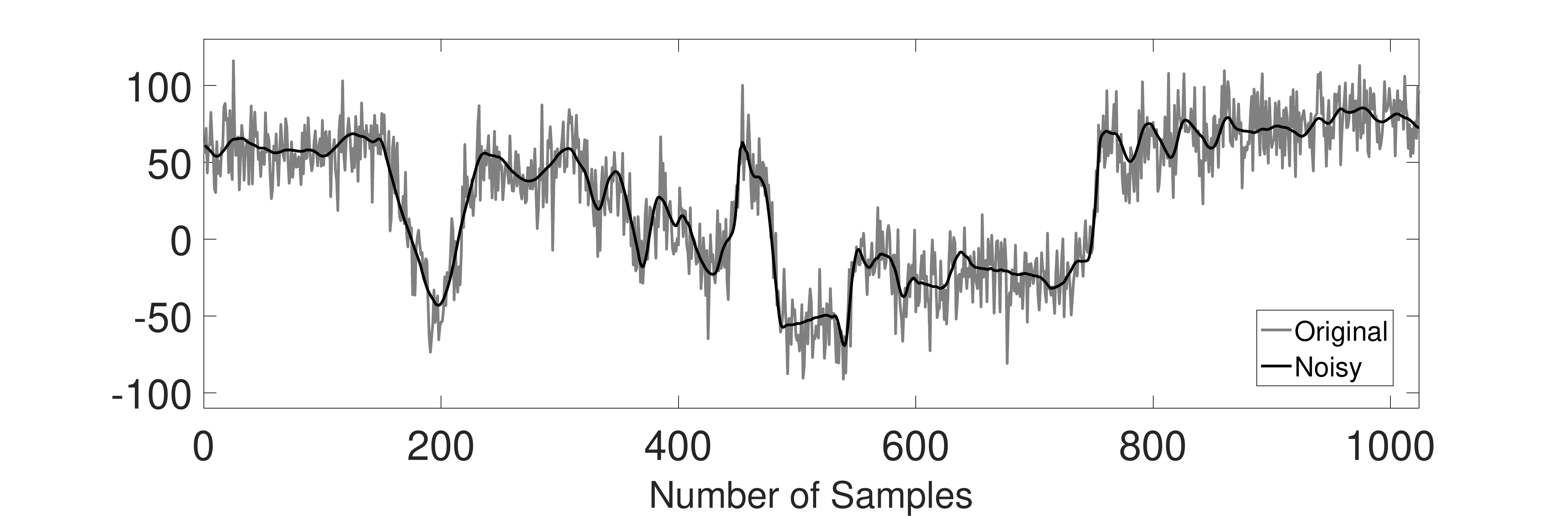}} \centerline{(a) \small{Noisy signal}} \end{minipage}
	\begin{minipage}[b]{1\linewidth} \centerline{\includegraphics[scale=0.22]{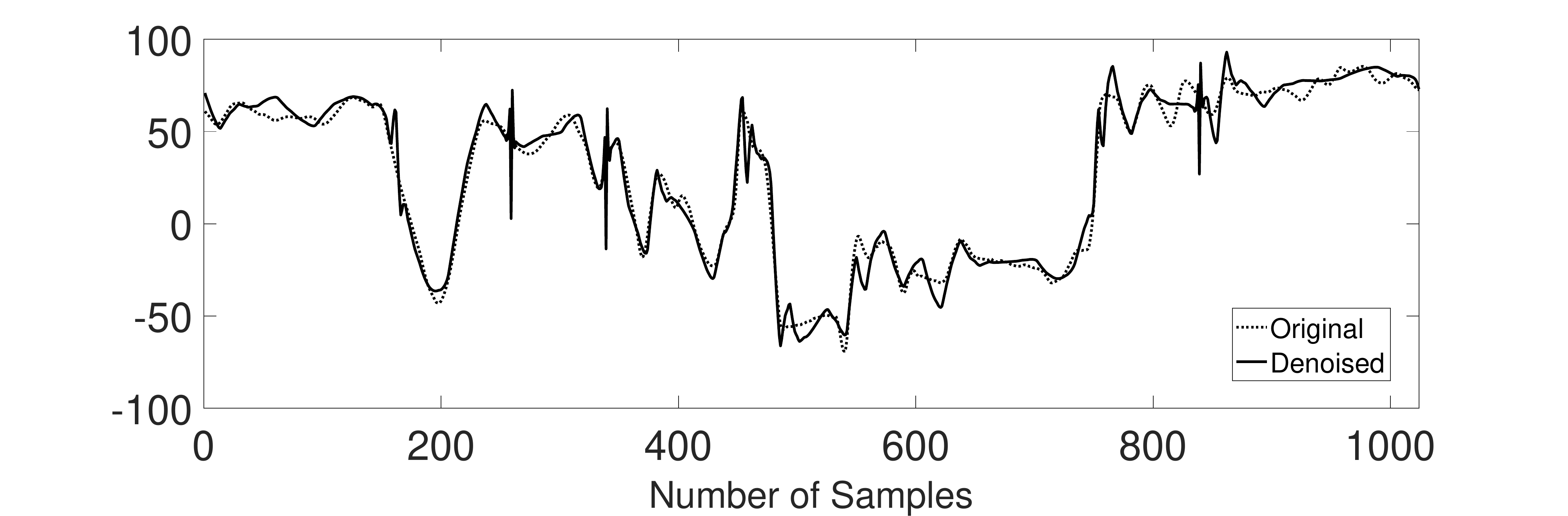}} \centerline{(b) \small{BLFDR}} \end{minipage}
	\begin{minipage}[b]{1\linewidth} \centerline{\includegraphics[scale=0.22]{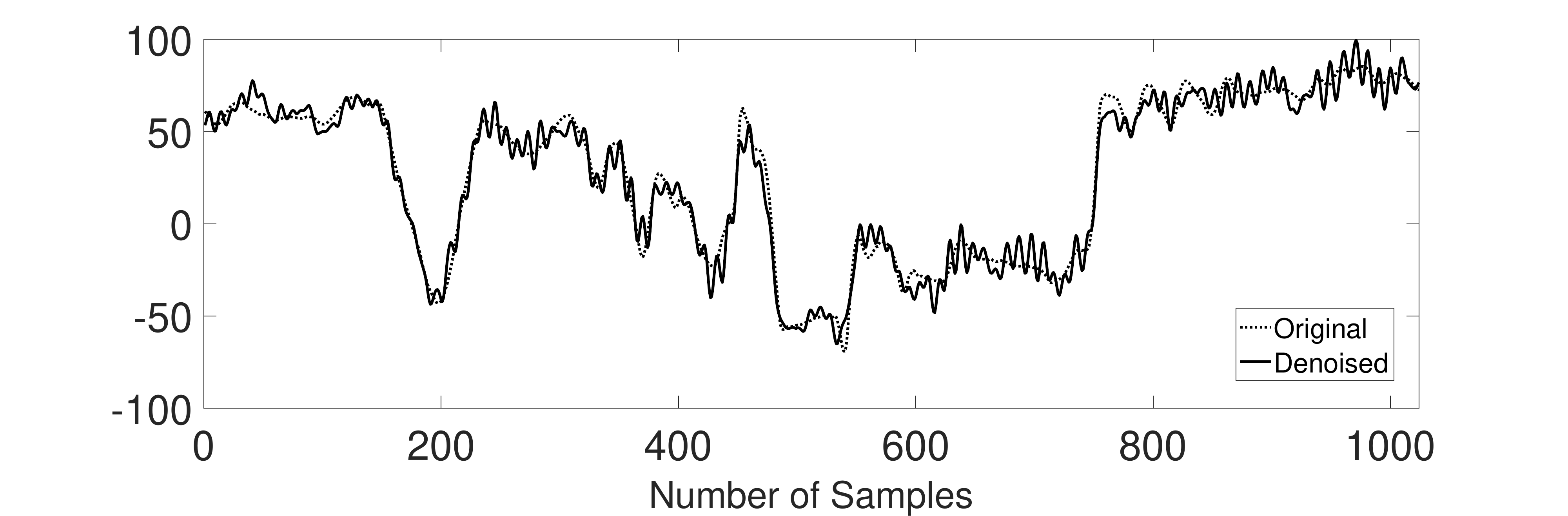}} \centerline{(c) \small{VMD-DFA}} \end{minipage}
	\begin{minipage}[b]{1\linewidth} \centerline{\includegraphics[scale=0.22]{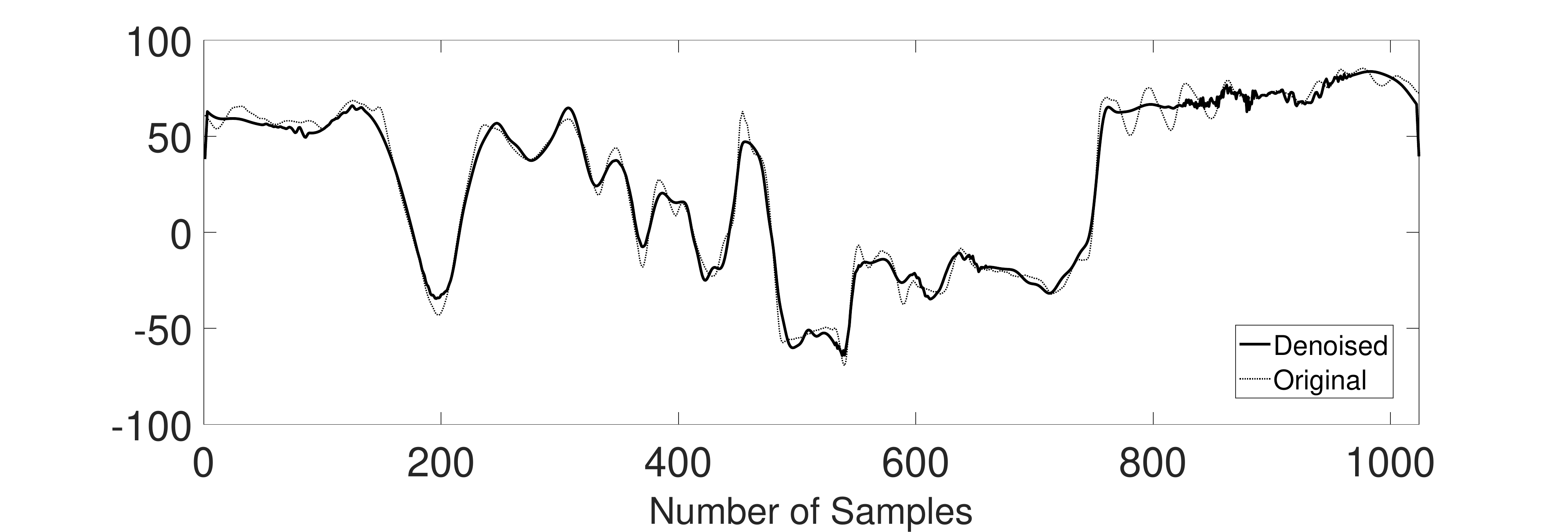}} \centerline{(d) \small{DT-GoF-NeighFilt}} \end{minipage}
	\begin{minipage}[b]{1\linewidth} \centerline{\includegraphics[scale=0.22]{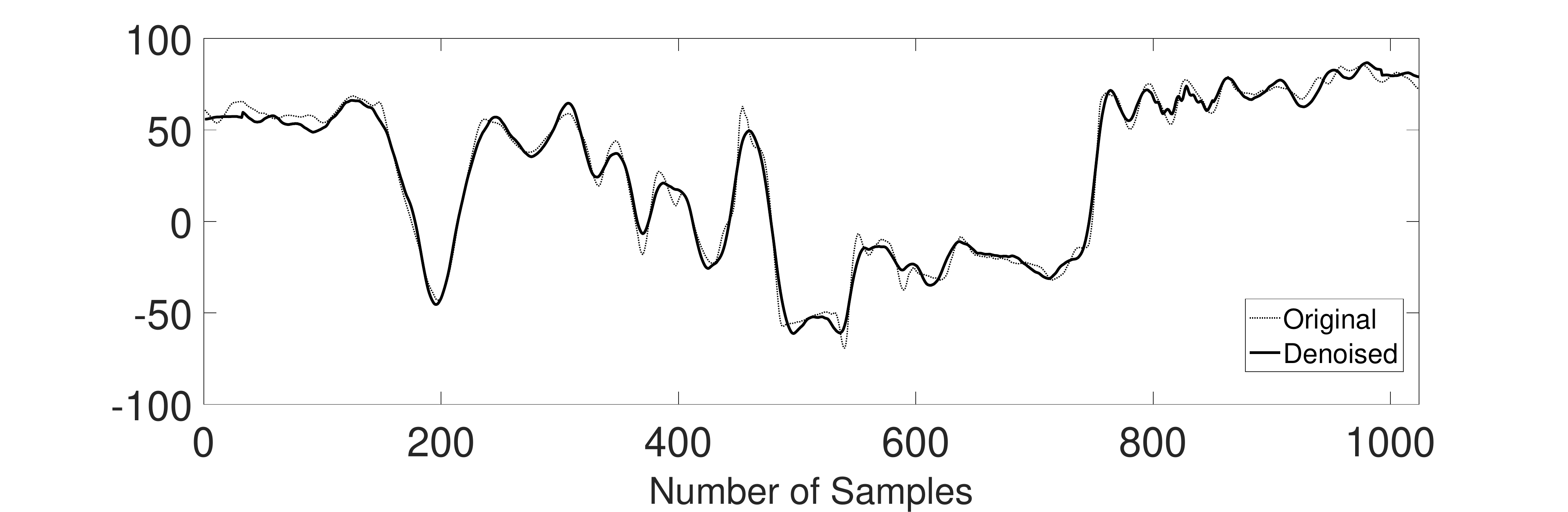}} \centerline{(e) \small{Prop. VMD-CVM}} \end{minipage}
	\caption{Noisy `Tai Chi' signal (a) and Denoised `Tai Chi' signals for various methods ((b) BLFDR, (c) EMD-IT, (d) DWT-GoF, (e) DT-GOF-NeighFilt, and (f) proposed VMD-CVM) for $SNR = 10$ dB.}
	\label{Ch1:DenoisedSet2}
\end{figure}
\subsection{Qualitative Performance Analysis}
The qualitative analysis demonstrates how closely the denoised signals from various methods resemble their corresponding true or noise-free signals. Generally, this is shown by plotting the denoised signals along with the original (noise-free) one that enables the reader to visualize how well the denoising methods extract signal-details from the noisy signal. To that end, we plot denoised `Bumps' and `Taichi' signals along with the original ones respectively in Fig. \ref{Ch1:DenoisedSet1} and Fig. \ref{Ch1:DenoisedSet2} whereby the corresponding noisy signals are also shown for comparison. We  compare the visual results of the proposed VMD-CVM method against the top comparative state of the art methods namely BLFDR, EMD-IT, VMD-DFA, GoF-DWT and DT-GOF-NeighFilt. Denoised signals were obtained by respectively denoising the noisy `Bumps' (shown in Fig. \ref{Ch1:DenoisedSet1}(a)) and noisy `Tai Chi' signal (shown in Fig. \ref{Ch1:DenoisedSet2}(a)) where noisy version of input SNR $=10$ dB is shown in gray while true signal is shown in dark black.

It is observed from the Fig. \ref{Ch1:DenoisedSet1} that the proposed VMD-CVM method yielded best estimate of the original signals. Observe that the denoised `Bumps' signals from DWT-GoF and DT-GOF-NeighFilt yield very close estimate of the original signals as can be seen from Fig. \ref{Ch1:DenoisedSet1} (c \& d) but both these methods suffer through artifacts. That overshadows their efficiency of extracting signal details when compared to the proposed VMD-CVM that yields an equally close estimate of the original signal but without artifacts, see \ref{Ch1:DenoisedSet1} (e). More visible spike artifacts are found in the denoised `Bumps' signals by EMD-IT, see from Fig. \ref{Ch1:DenoisedSet1}(b), which deteriorate the overall quality of the denoised signals when compared to the original ones.

The denoised `Tai Chi' signals from the comparative methods are plotted in Fig. \ref{Ch1:DenoisedSet2} (b)-(d) where it can be seen that BLFDR and VMD-DFA fail to recover the peaks and highly varying parts of the signal. That is owing to the complex structure of the 'Tai Chi' signal composed of subtle variations with high range of frequencies which pose a challenge to extract in presence of noise. A better estimate of the true signal is obtained by the DT-GOF-NeighFilt that largely recovers the variations while doing away with noise. Though, it fails to capture the subtle variations specially in the last half of the denoised signal, see Fig. \ref{Ch1:DenoisedSet2}(e). The best estimate of the `Tai Chi' signal is obtained by the proposed VMD-CVM that captures the subtle variations throughout the signal as can be observed from see Fig. \ref{Ch1:DenoisedSet2}(f). Apart from the sharp changes situated in the middle of this signal, the proposed method recovers all the details of the real `Tai Chi' signal demonstrating its effectiveness for complex real world signals.

Furthermore, the VMD-DFA yields exaggerated variations as artifacts in the aftermath of denoising process, see Fig. \ref{Ch1:DenoisedSet2} (c). That is owing to its partial reconstruction nature where relevant modes were selected to reconstruct the denoised signal and the presence of noise within the selected modes was ignored. Observe from Fig. \ref{Ch1:DenoisedSet2}(e) that the proposed approach does not suffer from this issue because our method performs thresholding on the selected relevant modes to reject the coefficients exhibition noise-like-statistics. Consequently, the reconstruction of the denoised signal based on cleansed thresholded BLIMFs successfully avoids the artifacts otherwise seen the results of the VMD-DFA method.

\section{Denoising ECG signal corrupted by sensor noise}
\label{EEGresults}
\begin{figure}[t]
	\begin{minipage}[b]{1\linewidth} \centerline{\includegraphics[scale=0.25]{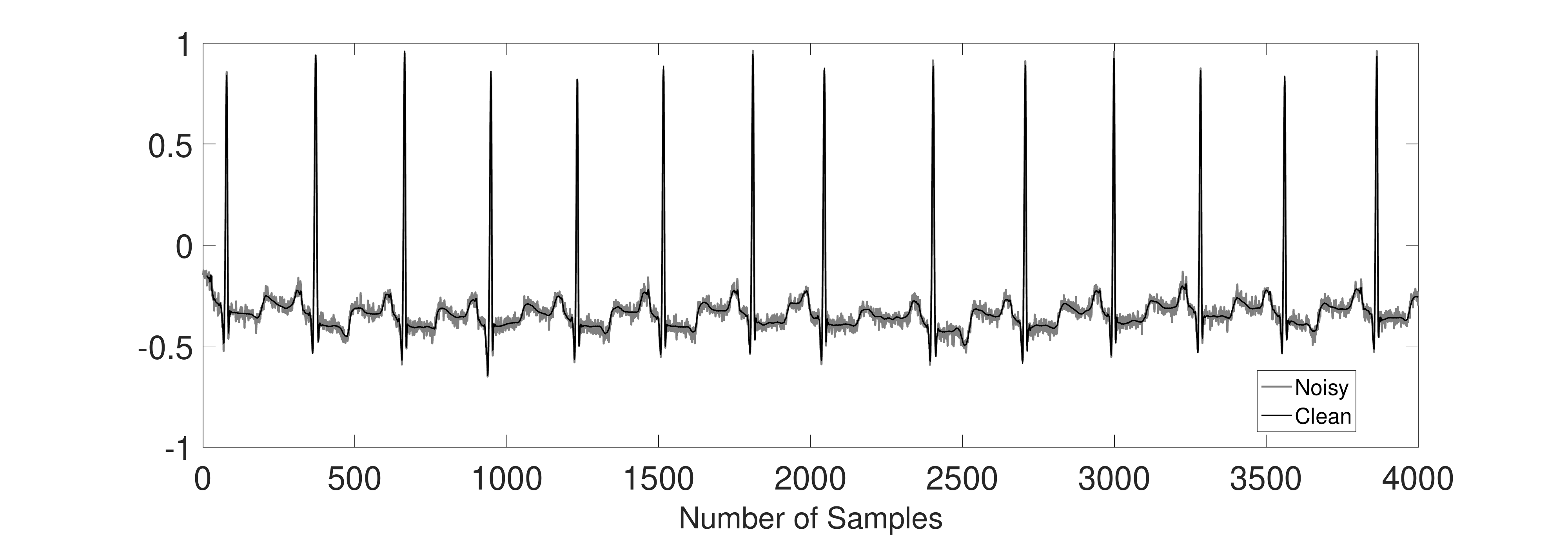}} \centerline{(b)} \end{minipage}
	\begin{minipage}[b]{1\linewidth} \centerline{\includegraphics[scale=0.25]{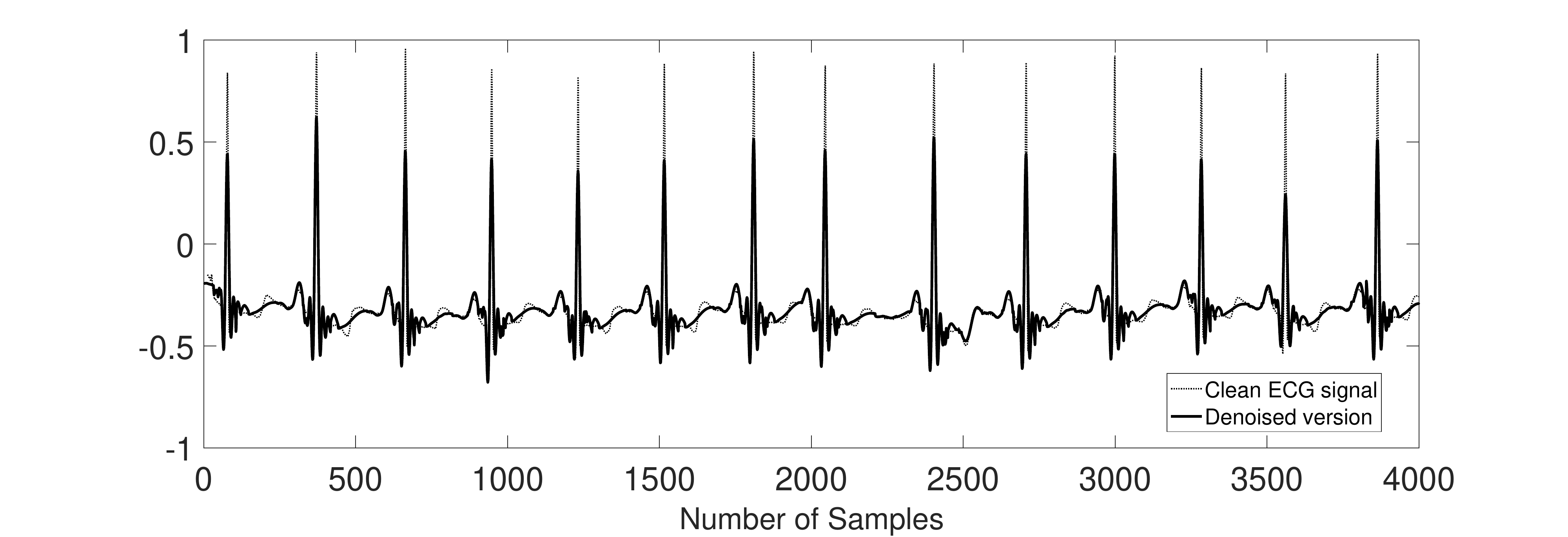}} \centerline{(b)} \end{minipage}
	\caption{Raw ECG signal along with its clean version obtained from \cite{tracey2012nonlocal} and its denoised version by the proposed VMD-CVM method.}
	\label{ECG}
\end{figure}

In this section, we present denoising results of the proposed method when applied to an ECG signal corrupted by the sensor noise. 
The raw ECG signal in this regard is taken from \cite{tracey2012nonlocal} that is corrupted by actual sensor that is typically modeled using the non-Gaussian distribution despite the presence of thermal noise due to electronic components that follows Gaussian distribution. As a result, the noise mostly obscures the useful information within the subtle variations of the ECG signal, observed from Fig. \ref{ECG} (a) where the noisy ECG signal in gray color along with a clean version of the raw ECG signal also available in \cite{tracey2012nonlocal} to be used as a ground truth.

To address this challenging problem we used the proposed method that can estimate the distribution of noise/ artifacts from within the noisy signal and subsequently use it to reject the noise. 
The resulting denoised signal is plotted in Fig. \ref{ECG} (b) (in dark black) where clean signal is also shown (in gray) in the background the denoised signal. Here, the effectiveness of our method is shown by demonstrating how closely the denoised version follows the clean ECG signal.
Evidently, from Fig. \ref{ECG} (b), the denoised version closely follows the clean signal in the background because it recovered important details including sharp peaks and slower variations. Despite the presence of the noise artifacts, observed near the sharp peaks, the overall quality of the recovered ECG remains intact verifying the efficacy of the proposed method in suppressing sensor noise while retaining the subtle variations which were previously hidden in the sensor noise.

\section{Conclusions}
\label{conclusion}
In this paper, we have addressed the problem of noise removal from the practical signals whereby the noise is considered to be governed by unknown probability distribution. We propose to exploit the desirable properties of VMD to estimate the EDF of noise from within the noisy signal. As stated earlier, VMD possesses ability to segregate the signal and noise in separate group of BLIMFs owing to its robustness to noise and mode mixing. First, we detect the group of BLIMFs predominantly composed of noise using CVM statistic followed by the empirical estimation of noise EDF from these rejected modes. Subsequently, the estimated distribution is used as a means to detect and reject the noise coefficients (the coefficients fitting the estimated noise EDF) from the remaining modes. The estimation of GoF of the reference noise EDF on the local segment has been performed by the CVM-GoF test.

The effectiveness of the proposed method has been demonstrated by comparing its performance against the state of art methods. It has been observed that the proposed method comprehensively outperformed the rest of the methods considered in this paper. In addition, the efficacy of the proposed method has also been demonstrated when addressing the problem of removal of sensor noise governed by some unknown distribution. For this purpose, we took the example of EEG signals (corrupted with sensor noise). It has been shown that the proposed method successfully removes the noise. The future prospects of this work may include its use in other practical applications for removal of noise before the signal processing part, e.g., denoising of Lidar signals, vibration signals from conditioning systems of heavy mechanical systems etc.

\bibliographystyle{IEEEtr}
\bibliography{manuscript}
\end{document}